# ATOMIC TRANSITION PROBABILITIES OF NEUTRAL CALCIUM[1]

E. A. Den Hartog[2], J. E. Lawler[2], C. Sneden[3], J. J. Cowan[4], I. U. Roederer[5,6] & J. Sobeck[7]

[2]Department of Physics, University of Wisconsin-Madison, 1150 University Ave, Madison, WI 53706; eadenhar@wisc.edu; jelawler@wisc.edu

[3]Department of Astronomy and McDonald Observatory, University of Texas, Austin, TX 78712; chris@verdi.as.utexas.edu

[4]Homer L. Dodge Department of Physics and Astronomy, University of Oklahoma, Norman, OK 73019; jjcowan1@ou.edu

[5]Department of Astronomy, University of Michigan, 1085 S. University Ave., Ann Arbor, MI 48109, iur@umich.edu

[6]Joint Institute for Nuclear Astrophysics – Center for the Evolution of the Elements (JINA-CEE)

[7]Univ. of Washington, Seattle, WA; jsobeck@uw.edu

## ORCIDS:

| | |
|---|---|
| E. A. Den Hartog: | 0000-0001-8582-0910 |
| J. E. Lawler: | 0000-0001-5579-9233 |
| C. Sneden: | 0000-0001-9566-3015 |
| J. J. Cowan | 0000-0002-6779-3813 |
| I. U. Roederer | 0000-0001-5107-8930 |
| J. S. Sobeck: | 0000-0002-4989-0353 |

[1] Based on observations made with the NASA/ESA Hubble Space Telescope, obtained at the Space Telescope Science Institute, which is operated by the Association of Universities for Research in Astronomy, Inc. under NASA contract NAS 5-26555. The data presented in this paper were obtained from the Barbara A. Mikulski Archive for Space Telescopes. Other data have been obtained from the European Southern Observatory Science Archive Facility.

ABSTRACT

The goals of this study are 1) to test the best theoretical transition probabilities for Ca I (a relatively light alkaline earth spectrum) from a modern ab initio calculation using configuration interaction plus many body perturbation theory against the best modern experimental transition probabilities, and 2) to produce as accurate and comprehensive a line list of Ca I transition probabilities as is currently possible based on this comparison. We report new Ca I radiative lifetime measurements from a laser-induced fluorescence (LIF) experiment and new emission branching fraction measurements from a 0.5 m focal length grating spectrometer with a detector array. We combine these data for upper levels that have both a new lifetime and new branching fractions to report log(*gf*)s for two multiplets consisting of nine transitions. Detailed comparisons are made between theory and experiment, including the measurements reported herein and a selected set of previously published experimental transition probabilities. We find that modern theory compares favorably to experimental measurements in most instances where such data exist. A final list of 202 recommended transition probabilities is presented, which covers lines of Ca I with wavelengths ranging from 2200 – 10,000 Å. These are mostly selected from theory, but are augmented with high quality experimental measurements from this work and from the literature. The recommended transition probabilities are used in a redetermination of the Ca abundance in the Sun and in the metal-poor star HD 84937.

# 1. INTRODUCTION

The story of early Galactic nucleosynthesis begins with detailed chemical compositions of low metallicity Milky Way halo stars. The record of the initial burst of massive star births, very short lives, and violent deaths is written in the abundance distributions of metal-poor stars. We must try to decode the abundance data to make any sense of how the Milky Way was born. But first we must determine the abundances accurately to have any hope of progressing beyond general guesses about what the early Galaxy did, when, how, and where.

Derivation of stellar abundances depends on many factors, from obtaining excellent high-resolution spectra, to constructing trustworthy stellar atmospheric models, to computing line absorption profiles with realistic radiative transfer techniques. But the efforts in these areas will be useless without high-quality line transition data. In particular, absolute atomic transition probabilities, or log($gf$)s, are critical to elemental abundance studies. Additionally, improved data on hyperfine and/or isotopic structure as well as improved energy levels are also vital, especially in cases where lines are saturated and/or blended.

In this paper we report improved transition probabilities for the first spectrum of the light "$\alpha$" element calcium[8]. Several $\alpha$ elements can be detected in very metal-poor stars: O, Mg, Si, S, and Ca. Their synthesis in massive stars has been understood for decades. However, these elements have generally not enjoyed recent comprehensive laboratory studies to the same degree as the Fe-group or neutron-capture elements. Ca is a relatively light alkaline earth, with small relativistic effects and with only two valence electrons. It is reasonable to hypothesize that for Ca I modern theory should be as accurate as modern experimental techniques. This hypothesis is tested in our work. There are a number of theoretical studies of Ca I in the literature. We choose the comprehensive work of Mills et al. (2017), hereafter M17, to make a detailed comparison to modern experiment.

---

[8]Formally an $\alpha$ element is one whose dominant isotope is composed of multiple $^{4}$He nuclei. The major natural isotopes of Ca (Z = 20) are $^{40}$Ca (96.94% in the Solar System), $^{42}$Ca (0.65%), $^{43}$Ca (0.14%), $^{44}$Ca, (2.09%), and $^{48}$Ca (0.19%); http://atom.kaeri.re.kr:81. For astrophysical purposes, Ca is pure $^{40}$Ca, making it the heaviest $\alpha$ element in the Periodic Table. Since the minor isotopes of Ca collectively contribute only 3% to the Ca elemental abundance, they will be undetectable in solar and stellar optical spectra. Finally, note that the first spectrum of calcium (i.e. Ca I) refers to the spectrum of the neutral species.

Section 2 of this paper contains a brief summary of recent theoretical calculations of Ca I, followed by a comparison of experimental radiative lifetime measurements to theoretical lifetimes derived from M17. New laser-induced fluorescence (LIF) lifetime measurements, accurate to 5%, are reported in §2. Published measurements from LIF and other laser-based measurements are also included in the comparison. Section 3 of this paper is a comparison of emission branching fractions (BFs) from our laboratory measurements and from the published literature to theoretical BFs from M17. In §3.4 we present new log(gf)s for nine transitions from two multiplets. Also discussed in §3 is our further development of calibration techniques based on a standard detector. A set of recommended log(*gf*)s for lines of Ca I is reported in §4. The recommended list of log(*gf*)s includes theoretical values from M17 and selected sets of measurements with some re-normalization. In §5 this list of recommended lines is used to re-determine the Ca abundance in the Sun and metal-poor star HD 84937. Finally, §6 includes a summary and some conclusions.

## 2. RADIATIVE LIFETIMES – THEORY AND EXPERIMENT

As noted above, neutral Ca is an alkaline earth with two valence electrons outside of a closed shell. These valence electrons yield an array of singlet and triplet levels. Spin-orbit splitting is a relativistic effect that is small in Ca I because of calcium's relatively light weight. This type of structure is also called LS or Russell-Saunders coupling. Figure 1 is a partial Grotrian diagram of Ca I which illustrates this singlet/triplet structure. (Note that in this figure the splitting between levels in a triplet term are exaggerated for the purposes of illustration.) Also shown in Figure 1 are the transition multiplets studied in the BF measurements described later in §3 (black lines connecting terms), as well as the multiplets studied in other experimental work (colored lines connecting terms) that we draw upon in §4 to generate our recommended list of log(*gf*)s. This figure should help the reader visualize the structure and related transitions in the detailed discussions below in the laboratory sections 3 and 4.

### *2.1 Recent Theoretical Studies of Ca I*

There have been a handful of quality theoretical calculations of Ca I in the last two decades. The Notre Dame team (e.g. Savukov & Johnson 2002) used the configuration interaction plus many body perturbation theory (CI+MBPT) method in an ab initio computation to find a small number of log($gf$)s, including transition probabilities for five spin-allowed resonance lines[9] and six spin-forbidden resonance lines. A larger calculation for neutral Ca was published by the Vanderbilt team (e.g. Froese-Fischer & Tachiev 2003) using the multi-configuration Hartree-Fock (MCHF) method which included all levels up to the $3d4p$ $^1F_3$. Transition probabilities were reported for ~24 spin-allowed lines and a smaller number of spin-forbidden lines. Most recently, M17 used the ab initio CI+MBPT method to compute transition probabilities for over 800 electric-dipole transitions of Ca I including most spin-allowed lines of interest and some spin-forbidden lines. Because of the comprehensiveness of this study, we have selected the work of M17 to test against our new measurements and other published measurements using modern methods.

The M17 data, which is included as supplemental material in that publication, is not in a user-friendly format as it requires proprietary software to easily access. Unfortunately the republication of the data by Yu & Derevianko (2018) has a few errors from misidentification of Rydberg levels. It has been recommended by an author that the supplemental material of M17 be used to resolve any discordance (A. Derevianko, private communication). There are no uncertainties quoted either in M17 or in the republication of the data by Yu & Derevianko. The latter publication does at least provide both the length and velocity forms of the calculation, as well as the percent difference between them (L-V) for many of the transitions studied. They found that the length form agreed better when compared to experiment and recommended that form, but found that for stronger lines the two forms agreed within a few percent. This L-V comparison can be used as a rough gauge of the theoretical uncertainties. Ca has attracted the attention of both theorists and experimentalists for decades and thus we omit comparisons to some of the older 208 publications on Ca I in the bibliography of the National Institute of Standards and Technology (NIST)

[9] The term resonance line refers to a transition that connects to the ground level.

Atomic Spectra Database (ASD) (Kramida et al. 2019). Results in many of the older publications have larger error bars than more recent studies. Our goal is to test the M17 theoretical results against the highest quality modern experimental results.

The M17 study was motivated by possible use of Ca in an optical frequency standard clock. Although microwave frequency clocks tied to a hyperfine transition of Cs are used as standard clocks today, it is anticipated that at some time those microwave frequency clocks will be replaced by optical frequency clocks. Very narrow optical transitions, when used for locking a clock oscillator, have the potential to make a standard clock with far greater (at least a million fold) accuracy and precision than can be achieved using a microwave transition. The same laser-cooled atom technology used in atomic clocks can be applied to transition probability measurements on certain resonance lines. Vogt et al. (2007) built on the work of Zinner et al. (2000) and Degenhardt et al. (2003) to measure the transition probability of the 4226.728 Å resonance line of Ca I, from the upper $4s4p\ ^1P^o_1$ to ground state $4s^2\ ^1S_0$ level, with an estimated accuracy and precision of ± 0.04%. This transition is indicated in Figure 1 with an orange line. (The transition probability of this Ca I spin-allowed resonance line was already known to a few percent accuracy and precision, which is adequate for most astrophysical research.) The laser-cooled atom technique is not broadly applicable to the many other optical and ultraviolet (UV) transitions of Ca I. However, the extensive theoretical work on >800 lines of Ca I in M17 is of great utility to astrophysics and to other fields.

### *2.2 Experimental Radiative Lifetimes*

Radiative lifetimes are measured using time-resolved LIF spectroscopy on neutral Ca atoms in a slow atomic beam produced in an electric discharge sputter source. This experiment has been successfully applied to many neutral and singly-ionized species throughout the periodic table over nearly four decades, and hence has been described many times in print. Here we will give a somewhat cursory description of the experimental technique. The reader is referred to Den Hartog et al. (2002) for a more detailed description of the experiment and an in-depth discussion of the various systematic effects associated with the measurements and how they are mitigated.

A gas phase sample of Ca atoms is produced by sputtering in a pulsed hollow cathode discharge operating with 30 mA DC and 5 – 10 A, 10 μs pulses in ~0.4 Torr argon. The hollow cathode is closed on the downstream end except for a 1 mm hole through which the beam exits. The cathode is made of steel and is typically lined with a high-purity thin sheet of whatever metal is being studied. Calcium, however, is very reactive and oxidizes so readily that it is not generally available in sheet form. Instead, we had an 80% Al - 20% Ca alloy disk manufactured to line the bottom of the cathode and used 1100 aluminum shim stock to line the vertical walls of the cathode. The alloy was chosen as it is much less reactive than pure Ca and therefore easier to sputter off the oxide layer during cathode conditioning. The Ca atoms and ions exit the cathode through the 1 mm hole, entrained in a flow of argon gas, into a scattering chamber which is held at $\sim 1\times10^{-4}$ Torr. The atomic beam environment has been tested and repeatedly shown to be free from effects related to collisional depopulation and optical depth. This beam of Ca atoms is slow ($\sim 5\times10^{4}$ cm/s) and weakly collimated and has a mix of ground and metastable level populations. In neutral Ca, the ground level is a $4s^2\ {}^1S_0$ and the lowest metastable levels are in the $4s4p\ {}^3P^o$ term at ~15,000 $cm^{-1}$ (see Figure 1). Using these populations as lower levels, we have access to odd-parity ${}^1P^o$ terms as well as even parity ${}^3S$, ${}^3P$ and ${}^3D$ terms using single-step laser excitation. The even-parity $4s4d\ {}^1D_2$ metastable level at 21849 $cm^{-1}$ is too weakly populated in our beam to make use of as the lower level of a single-step excitation.

A laser beam intersects the atomic beam at right angles 1 cm below the bottom of the cathode and is used to selectively excite the level being studied. Selective excitation is an important advantage of laser-based lifetime measurements. Older techniques that relied on nonselective electron beam excitation were prone to systematic error due to cascade from higher lying levels. The laser used in this study is a pulsed dye laser pumped with a nitrogen laser. It has a bandwidth of ~0.2 $cm^{-1}$ and is tunable from ~2050 – 7000 Å using a variety of dyes and frequency doubling crystals. The laser pulse is ~3 ns duration and terminates completely within a few nanoseconds of peak intensity. This abrupt termination makes it possible to record the fluorescence decay free from laser interaction. The laser is triggered 20 – 30 μs after the peak of the discharge current, allowing transit time for the atoms to reach the beam interaction region. Light from the laser is polarized along the axis of the atomic beam resulting in the possibility of Zeeman quantum beats.

This is a phenomenon which arises because the excitation from the polarized laser leaves the atoms in a dipole-aligned state. These dipoles then precess about the earth's magnetic field while radiating, sweeping the non-uniform dipole radiation pattern through the direction of view, resulting in an oscillation of the fluorescence signal. To mitigate this effect, the field is zeroed to within ±0.02 Gauss in the viewing volume using a set of Helmholtz coils. When long lifetimes (>300 ns) are measured a high magnetic field (30 Gauss) is imposed along the laser axis so that the precession is very fast relative to the time scale of the measurement, and the oscillations in the fluorescence average away. The longest Ca I lifetime measured in this study is <50 ns, so the high field was not required.

Fluorescence is collected in a direction perpendicular to the plane defined by the laser and atom beams. A pair of fused silica lenses are used to image the beam interaction region with unity magnification onto the photocathode of a RCA 1P28A photomultiplier tube (PMT). Optical filters, either broadband colored glass filters or narrowband multilayer dielectric filters, are often placed between the two lenses where the fluorescence is roughly collimated in order to block scattered laser light as well as cascade radiation from lower lying levels. Occasionally, for the measurement of long lifetimes, it is also necessary to place a cylindrical lens between the two fluorescence collection lenses to defocus the light at the PMT. This is to mitigate the flight-out-of-view effect, in which the image of atoms fluorescing later in the decay has moved to a region of lower sensitivity on the PMT photocathode, resulting in a perceived shortening of the lifetime. The flight-out-of-view effect only becomes a problem for lifetimes longer than 300 ns for neutral atoms and 100 ns for ions, which move somewhat faster in the beam, and thus was not an issue in our measurement of the Ca I lifetimes in this study.

The PMT has a rise time of ~1.7 ns and the electrode chain is carefully wired for low inductance to minimize ringing. The PMT signal is put through a high bandwidth 60 ns delay line and into a Tektronix SCD1000 transient digitizer for signal recording. The bandwidth and fidelity of the detection electronics is such that only lifetimes below 3 ns show any systematic distortion. The shortest lifetime we have measured in Ca I is 4.6 ns, well above this limit. Another systematic arising from the PMT is caused by an after-pulse signal. After-pulsing arises from an imperfect vacuum within the PMT. As the fast electron avalanche proceeds from photocathode to anode, some ionization of residual gas in the PMT occurs. These ions make their way toward the cathode, but on a much longer time scale given the much higher mass of the ions compared to that of the electrons. The result is a very weak pulse ~150 ns after the fast pulse and spread over 10's of nanoseconds. This effect results in a lengthening of lifetimes in the 100 – 150 ns range by a few percent. This effect can be reduced by periodically introducing a light leak into the optical system for a number of hours resulting in a 1 μA DC PMT anode current that causes the ionized gas to be buried in the cathode. This "degassing" of the PMT does not eliminate the effect altogether but reduces it substantially.

The desired transition wavelength is found by first coarse tuning the laser wavelength to within ~0.5 Å as measured by a 0.5 m focal length monochromator. Coarse tuning is accomplished by tipping the grating, which is the tuning element of the dye laser. An LIF spectrum is then recorded over a 5 – 10 Å range by slowly changing the pressure of nitrogen in an enclosed volume surrounding the grating. This "pressure scanning" gives very fine and reproducible control of the laser wavelength. Data acquisition involves recording an average of 640 fluorescence decays, starting only after the laser has completely terminated and for a time span equivalent to approximately three lifetimes. The laser is then tuned off the transition and an average of 640 backgrounds is recorded. The lifetime is determined, both for early time (first half of the trace) and late time (second half of the trace) by doing a least-squares fit to a single exponential on the difference between the signal and background records. Comparison of the early and late lifetimes is a quick and sensitive way to determine if the decay is a clean exponential or if it has some distortion from a systematic effect, such as cascade radiation through lower levels, that needs further study. A set of five of these acquisitions and analyses comprise one lifetime determination. The lifetime is determined twice for each upper level using a different laser transition whenever possible. This redundancy helps to ensure that the transitions are identified correctly in the experiment, that they are correctly classified to the level and that neither transition is blended.

Systematic effects in the experiment are well understood and controlled at the ±5% level, which is our quoted uncertainty in most cases. To ensure that our experiment is operating reproducibly and our measurements lie within the stated uncertainties, we routinely measure a set of "benchmark" lifetimes. These are lifetimes that are accurately known from other sources, and have been determined either from theory or from an experiment with different and smaller systematic uncertainties than in our experiment. To cover the range of lifetimes measured for Ca I we have measured the following benchmark lifetimes: $2^2P_{1/2,3/2}$ levels of $Ca^+$ at 6.904(26) ns (Hettrich et al. 2015) and 6.639(42) ns (Meir et al. 2020), respectively; $2^2P_{3/2}$ level of $Be^+$ at 8.8519(8) ns (variational method calculation, Yan et al. 1998); the $3^2P_{3/2}$ level of neutral Na at 16.23(1) ns (NIST critical compilation of Kelleher & Podobedova 2008; uncertainty of ±0.1% at 90% confidence level); $4p'[1/2]_1$ level of neutral Ar at 27.85(7) ns (beam-gas-laser-spectroscopy, Volz & Schmoranzer 1998). We also note that our remeasurement of the Ca 4*s*4*p* $^1P_1$ lifetime similarly acts as a "benchmark" end-to-end check of our experiment, as this lifetime is known to very high accuracy and precision from other sources (e.g. Vogt et al. 2007).

The results of our radiative lifetime measurements are given in Table 1 along with all other experimental lifetimes measured using LIF or other modern laser-based methods. In addition, we compare to lifetimes determined from the theoretical transition probabilities of M17. We do not include comparisons to older experimental methods that involved non-selective excitation. Configurations, terms and level energies in Table 1 and in subsequent text, tables and figures are taken from the NIST ASD (Kramida et al. 2019). Air wavelengths used throughout the text and tables are calculated using these energy levels and the index of air from Peck & Reeder (1972).

## 3. EMISSION BRANCHING FRACTIONS AND log(*gf*) VALUES

The combination of radiative lifetimes from LIF measurements with emission BFs from Fourier transform spectrometers (FTSs) or other high-resolution spectrometers has proven to be an efficient and accurate method for measuring *A*-values and log(*gf*)s (e.g. Lawler et al. 2009). The radiative lifetime of an upper level *u* provides the absolute scale when converting the BFs of transitions connected to that level to *A*-values. The BF for a transition between *u* and a lower level *l* is the ratio of its *A*-value to the sum of the *A*-values for all transitions associated with *u*, which is the inverse of the radiative lifetime, $\tau_u$. This can also be expressed as the ratio of relative emission intensities I (in any units proportional to photons/time) for these transitions:

$$BF_{ul} = \frac{A_{ul}}{\sum_l A_{ul}} = A_{ul}\tau_u = \frac{I_{ul}}{\sum_l I_{ul}}. \quad (1)$$

BFs, by definition, sum to unity, or near unity if a few residual weak lines are not measured. It is therefore important when measuring BFs to account for all possible decay paths from an upper level so that the normalization is correct. If some significant transitions are omitted then one has a branching ratio (BR) rather than a BF.

### 3.1 *BFs of Triplet Multiplets*

Triplet multiplets – the transitions connecting the levels in an upper triplet term to those in a lower triplet term - typically are spread over one, or at most a few percent fractional change in wavelength due to the close spacing of levels in each term. The emission BF measurements on triplet multiplets reported herein were made using a Jarrell-Ash 0.5 m focal length grating spectrometer equipped with a 1180 groove/mm diffraction grating blazed for 3000 Å. The grating spectrometer was equipped with a Si photodiode array having 1024 pixels, each 25 microns wide. An internal relative radiometric calibration technique employing Ar I and Ar II lines (Whaling et al. 1993) has been used heavily by our group in the analysis of FTS data from other species. This technique is desirable when calibrating spectra over wide wavelength ranges because it captures wavelength dependent effects such as window transmission and internal reflections in the lamp source. The minimal wavelength spread of the multiplets studied here negates the need for this method, and an external calibration is used herein. The spectrometer and detector array are calibrated over the small wavelength ranges needed using a NIST traceable tungsten-quartz-halogen (WQH) standard lamp operated at 6.5 amps.

Table 2 reports our emission BF measurements on a triplet multiplet from the upper 4*s*5*s* $^3S_1$ level to the 4*s*4*p* $^3P^o$ term. These are true BFs with residuals << 0.01. Comparison of our measurements to BF measurements from Aldenius et al. (2009), to theoretical BFs from M17, and to pure LS (see e.g. Appendix I of Cowan 1981 for the tabulation of LS intensities from perturbation theory) are included in Table 2.

Emission BFs for this particular multiplet, as measured in the present study and by Aldenius et al., are found to be a nearly pure triplet that follows predicted LS BFs to a good approximation. This can be seen in panel a of Figure 2, where we plot the difference of log(BF) between our results and those of M17 versus the M17 BFs. In panel a, we also make the same comparison between LS BFs and those of M17. In this plot as well as in panels b-e, the horizontal line at 0.00 indicates perfect agreement with the M17 calculations. The same multiplet was studied earlier using absorption spectroscopy by Smith & O'Neill (1975) who also found nearly pure LS results. Similarly, Table 3 reports our emission BF measurements on a triplet multiplet from the upper $4p^2\ ^3$P term to the $4s4p\ ^3$P$^o$ term. Again, these are true BFs with residuals << 0.01. A comparison of our measurements to theoretical BFs from M17 and pure LS BFs are given in Table 3 and plotted in panel b of Figure 2. As with the previous multiplet, both experiment and theory show extremely good agreement with simple LS coupling theory.

Results presented in Table 2 and 3, and panels a and b of Figure 2, are for even-parity upper triplet terms. It is reasonable to check a few odd-parity upper triplet terms. Table 4 reports our emission BF measurements on a triplet multiplet from the upper $3d4p\ ^3$D$^o$ term to the $3d4s\ ^3$D term. These are true BFs with residuals < 0.01 as indicated by the BFs from M17. Table 4 includes a comparison of our measurements to theoretical BFs from M17 and to pure LS BFs, and these comparisons are presented also in Figure 2c. As can be seen in the figure, this multiplet also follows LS BFs to a good approximation, although not quite as closely as the previous two even-parity multiplets. This multiplet was studied earlier using absorption spectroscopy by Smith & Raggett (1981) who also found nearly pure LS results. Nearby triplet levels of the same parity and J of the upper levels of this multiplet can "repel" triplet levels of the upper term and help explain the small deviations between pure LS BFs and those calculated by M17. Finally, Table 5 reports our emission BF measurements on a triplet multiplet from the upper $3d4p\ ^3$P$^o$ term to the $3d4s\ ^3$D term. For measurement of the weakest line of the multiplet, an echelle grating was substituted for the first order grating in the Jarrell-Ash 0.5 m spectrometer in order to increase the instrument resolving power so as to isolate the line from nearby stronger lines. These are true BFs with residuals of ~0.008 to 0.009 as indicated by the BFs from M17. Table 5 includes a comparison of our measurements to theoretical

BFs from M17 and to pure LS BFs, and these comparisons are plotted in Figure 2d. This multiplet also follows LS BFs to a good approximation. Like the multiplet of Table 4 this multiplet was studied earlier by Smith & Raggett (1981) who also found nearly pure LS results using absorption spectroscopy.

### 3.2 *BFs Involving Mixed Levels*

The data in tables 2 – 5 confirm that LS coupling is quite good in neutral Ca, which is not surprising because Ca is a light alkaline earth element. The breakdown of LS coupling is expected in sufficiently high Rydberg levels where relativistic effects can overcome Coulomb interaction of the valence electrons. Of course, the breakdown of LS coupling occurs in low-lying levels of heavier atoms. There **are** only very small deviations from LS in the multiplets of Table 2 through 5 and these deviations are near the limit of or below our detection threshold. There are a few stronger breakdowns of LS coupling in low-lying levels of neutral Ca and it is thus appropriate to test the M17 log(*gf*)s in at least one case where there is a detectable breakdown. The two good quantum numbers which govern LS breakdowns are J, the total electronic angular momentum, and parity determined by the configuration. The $^1D^o_2$ level at 35835.413 cm$^{-1}$ and the $^3F^o_2$ level at 35739.454 cm$^{-1}$, both of the upper 3*d*4*p* configuration, are significantly mixed due in part to their small energy separation. In Table 6 we report BF measurements for the decay of the 3*d*4*p* $^1D^o_2$ level and the three 3*d*4*p* $^3F^o_{2,3,4}$ levels to the 3*d*4*s* $^1D_2$ and 3*d*4*s* $^3D_{1,2,3}$ lower levels. Although not all of the lines from upper J = 2 levels fit on a single photodiode array exposure, the lines are sufficiently close together in wavelength for a satisfactory relative radiometric calibration using our WQH standard lamp. These are true BFs with residuals < 0.005. A comparison of our measurements to theoretical BFs from M17 is included. Although pure LS BFs are also included for comparison, those BFs omit the $J_{upp} = 2$ level mixing. This LS breakdown yields mixed levels and leads to violations of the $\Delta S = 0$ spin-selection rule of LS coupling for the J = 2 upper levels. A comparison between both our measured BFs and LS BFs and those of M17 are plotted in Figure 2e. The wide deviation from LS is apparent in this figure, as is the good agreement between our emission BF measurements and the calculations of M17. Smith and Raggett (1981) also measured log(*gf*)s for these multiplets using absorption spectroscopy.

### 3.3 *BRs of Singlet Transitions*

On the singlet side, BF measurements often involve widely separated wavelengths, significantly increasing the difficulty of the measurement. The BFs for strong lines tend to depend on radial wave functions because different configurations are involved. One such case is the resonance line at 2398.559 Å connecting to the upper 4*s*6*p* $^1P^o_1$ level at 41679.008 $cm^{-1}$. This case stood out when the M17 results were compared to hook measurements by Ovstrovskii & Penkin (1961) and Parkinson et al. (1976). M17 reported a log(*gf*) smaller than the hook experiments by 0.26 and 0.23 dex, respectively. On a more positive note, the spin-forbidden resonance line connecting to the 4*s*4*p* $^3P_1$ level at an air wavelength 6572.779 Å, has a log(*gf*) of -4.274 in M17. This value is in good agreement with the log(*gf*) = -4.24 reported by Drozdowski et al. (1997) based on a LIF experiment and with the log(gf) = -4.32 reported by Parkinson et al. (1976). The spin allowed resonance line at 4226.728 Å connecting to the 4*s*4*p* $^1P_1$ level has log(*gf*) of 0.242 in M17, of 0.243 in Parkinson et al. (1976), and of 0.23884(9) in the laser cooled atom experiment by Vogt et al. (2007).

The dominant line from the 4*s*6*p* $^1P^o_1$ level at 41679.008 $cm^{-1}$ connects to the 3*d*4*s* $^1D_2$ level at 21849.634 $cm^{-1}$ in the optical at 5041.618 Å. This transition has BF = 0.658 in the data from M17. The UV resonance line at 2398.559 Å has a BF = 0.254 in the data from M17. These two transitions are indicated with black lines in Figure 1. The BR = 0.254 / 0.658 = 0.386 is an attractive test of the M17 theoretical transition probabilities. This is not a trivial BR measurement because it involves bridging a relative radiometric calibration from the optical to the UV as discussed by Lawler & Den Hartog (2019). The UV wavelength of interest is significantly beyond the calibration limit of the Ar I and Ar II method (Whaling et al. 1993). In such a case, a calibration based on a standard detector is advantageous. Our measurement of the UV line with respect to the visible line is based on a NIST calibrated Si photodiode (PD) as used by Lawler & Den Hartog. Radiation from a line, or preferably continuum, source is measured using the calibrated PD and measured using the spectrometer plus detector array to transfer the PD calibration to the spectrometer plus detector array. The very stable Hg pen lamp which was used by Lawler & Den Hartog is replaced by a Xe arc lamp for the deep UV in this work. This lamp has a substantial amount of flicker, but a method was found to overcome this with signal averaging. The Si diode impedance is too low for introducing a multi-second averaging with a capacitor. An unrealistically large capacitance is required. However if the signal from the Si diode is measured with an electrometer, then a ~5 sec averaging can be easily introduced between the output of the electrometer and the input of a digital multimeter with a high input impedance. In the earlier study by Lawler & Den Hartog, the radial temperature variation of the Hg pen lamp was overcome by rotating the lamp so that the radial variation lay along the length of the entrance slit of the spectrometer. Arc lamps, however, are generally run in a vertical orientation to avoid "arching" of the discharge, so rotating the lamp itself was inadvisable. Instead, we rotated the image of the lamp. This can be done either with a prism or with a pair of mirrors. Multi-layer dielectric (MLD) filters with a 100 Å pass band were used to isolate a wavelength interval from the Xe arc lamp for measurement with the NIST calibrated Si diode. In spectral regions where the spectrometer plus detector array calibration is relatively flat, as shown in Fig. 2 of Lawler & Den Hartog (2019), only a few MLD filters are needed. In the UV where the relative radiometric calibration of the spectrometer plus photodiode is steep, more MLD filters are needed. Lastly we should mention attempts to use a small "in line" 0.2 m focal length grating monochromator as a prefilter to calibrate the Jarrell-Ash 0.5 m focal length spectrometer. This initially seemed attractive because it could reduce the number of MLD filters needed. However, the polarization and angular variations from the combination of two diffraction grating instruments were so troublesome that we resorted to MLD filters, including one centered at 2398 Å. This wavelength is near the lower limit of a calibration using a Xe arc lamp and calibrated Si PD. The power transmitted by the MLD filter needs to be sufficient for a high signal-to-noise (S/N) measurement using the Si diode.

Our final measurement is BR = 1.043 ± 10% for the UV over optical 2398.559 Å / 5041.618 Å ratio. The reader may notice that only a BR measurement is reported here because the optical and UV lines in the BR are connected to the ${}^1P^o_1$ upper level at 41679.008 cm$^{-1}$ which has residual decays of 0.09, according to M17. Because we have determined a BR rather than a complete set of BFs for all transitions to the upper level, we cannot determine a log(*gf*) to directly address the discrepancy between the hook measurements and M17 for the resonance transition. Our measured BR is much larger than the BR = 0.386 computed by M17. Further evidence that it is the resonance line that is off in the M17 calculation rather than the optical transition from this pair can be seen when the radiative lifetime for the level is compared to our measured radiative lifetime (see Table 1.) We measured 20.6 ± 1.0 ns whereas that calculated by M17 is 27.7 ns. However, if one increases their resonance line strength to give a BR commensurate with our measured BR but leaving all other A-values as calculated by them, the lifetime of the level would be 19.3 ns which is in much better agreement with our measured lifetime. Alternatively, if the M17 A-value for the 5041 Å line was decreased to match our measured BR, then the calculated and measured lifetimes would only get further apart. Although we initially planned to emphasize more recent measurements, it is clear that Parkinson et al. (1976) considered the relative hook measurements of Ostrovskii & Penkin (1961) to be of exceptional quality. Parkinson et al. used the then well-known log(*gf*) = 0.243 of the spin-allowed resonance line at 4227 Å to put the older relative measurements on a reliable absolute scale. We are thus recommending the log(*gf*) of Ovstrovskii & Penkin for the 2398 and 2721 Å resonance lines, and Parkinson et al. log(*gf*)s for other resonance lines except the 4227 Å line, for which we recommend the high precision measurement of Vogt et al. (2007). It is worth mentioning again, however, that the log(*gf*)s from the calculations by M17 agree with the measurements of Parkinson et al. for the 2200 Å, 2275 Å, and 6572 Å resonance lines and with the measurement of Vogt et al. for the 4227 Å spin-allowed resonance line.

### *3.4 log(gf) Values from Lifetimes and Branching Fractions*

Our lifetime measurements include the upper levels of transitions from the 4*s*5*s* $^3$S, 4*s*4*d* $^3$D, and 4p$^2$ $^3$P terms used by Ueda et al. (1982, 1983) as reference transitions for their hook measurements. These measurements and their renormalization are discussed in §4. The BF's reported in Tables 2 and 3 for two of these terms, 4*p*$^2$ $^3$P and 4*s*5*s* $^3S_1$, are combined with the radiative lifetimes for those upper levels from Table 1 to produce *A*-values and log(*gf*)s for nine transitions. These are presented in Table 7 along with log(*gf*)s from M17 and from Aldenius et al. (2009). For these multiplets we see excellent agreement with M17, with their log(*gf*)s agreeing with those of this study within 0.025 dex for all nine transitions. The agreement with the experimental measurements of Aldenius et al. is not quite as good, with their log(*gf*)s being 0.04 – 0.06 dex smaller than our result for the three lines in common. Most of this difference is due to their lifetime measurement for the 4*s*5*s* $^3S_1$ being 10% longer than our result.

## 4. RECOMMENDED log(*gf*)s

This section establishes a set of recommended log(*gf*)s for lines of Ca I ranging in wavelength from 2200 – 10,000 Å. Lines with wavelengths ≤ 10,000 Å are compatible with Si CCD detector technology and are included herein. The development of HgCdTe detector arrays is opening the Infrared (IR) but Laboratory Astrophysics has not caught up with IR observations. We are augmenting the theoretical log(*gf*)s of M17 with sets of measurements including published measurements and including some of ours described above. There is an augmented set of log(*gf*)s in the supplemental material of M17, but our set is updated from that set.

In the preceding section we discussed our measurement of the BR for the 2398.559 Å resonance line. This measurement suggests that log(*gf*)s from the experimental hook measurements are more reliable than M17 for resonance lines, and are adopted herein. This affects only five transitions, indicated in Figure 1 with blue lines, and we note that experimental log(*gf*)s for three of the five resonance lines are in agreement, within uncertainties, with those computed by M17. The discordance in log(*gf*) values for the weak resonance line at 2721.644 Å deserves some additional study if it is used for abundance measurements. There is no doubt that the log(*gf*) = 0.23884(9) of the resonance line at 4226.7276 Å measured by Vogt et al. (2007), indicated with an orange line in Figure 1, is superior to other measurements and is thus included herein. Parkinson et al. normalized their log(*gf*)s using log(*gf*) = 0.243 for this transition based on an earlier measurement by Smith & Liszt (1971). Parkinson et al. reported other log(*gf*) measurements to 0.01 dex. The normalization of Parkinson et al. (1976) is offset by approximately +0.004 dex or +0.96% from the precise and accurate measurement on the 4227 Å line by Vogt et al. (2007). Without log(*gf*)s reported to 0.001 dex it is not possible to make such a small renormalization. Uncertainties on the log(*gf*)s reported by Parkinson et al. are ± 0.06 dex except for the weak line at 2721.644 Å which is a bit higher. Parkinson et al. (1976) indicates that Ostrovskii & Penkin (1961) have a smaller error bar on the resonance lines at 2398 Å and 2722 Å than their newer measurements. Ostrovskii & Penkin's relative hook measurements were put on an absolute scale using the log(*gf*) of the spin-allowed resonance line at 4227 Å as discussed above, and we recommend theirlog(*gf*)s for these two lines. It should be said, however, that the two sets of hook measurements agree within uncertainties, lying only 0.03 dex apart for the 2398 Å line and 0.01 dex apart for the weak 2722 Å line.

The hook measurements by Ueda et al. (1982) and Ueda et al. (1983) are relative *gf* measurements normalized to published radiative lifetime measurements. Their systematic uncertainty is at least 10% from the absolute scale for their relative hook measurements. This 10% uncertainty can easily be reduced using our lifetime measurements and the M17 or LS branching ratios inside a multiplet. The fact that Ueda et al. reported relative *gf*s to 0.001 can be used to test M17 calculations of radial matrix elements for several multiplets and can be used to derive improved recommended log(*gf*)s. Our own experimental measurements confirm that Russell-Saunders or LS coupling is quite strong inside most multiplets of Ca I connecting low-lying levels. Based on our BF measurements in §3, we are recommending final transition probabilities that preserve the excellent multiplet coupling of M17 and use Ueda et al. (1982, 1983) measurements to refine radial matrix elements for each of four multiplets. These multiplets are indicated by red lines in Figure 1.

Ueda et al. (1982) measurements have lines in common with M17 on the UV and blue multiplets connecting the upper $4s5d\ ^3$D, $4s6s\ ^3$S, and $4p^2\ ^3$P terms to the lower $4s4p\ ^3$P term, with reference measurements on the two longest wavelength multiplets connecting upper $4s4d\ ^3$D and $4s5s\ ^3$S terms to the lower $4s4p\ ^3$P term. The first two upper terms have non-negligible IR residuals. This means that the UV and blue multiplet BFs sum to appreciably less than 1 for upper levels of those two multiplets. The third, $4p^2\ ^3$P and the two upper reference terms have negligible residuals. The two reference terms have new lifetime measurements as discussed in §2. We make small normalization corrections of +0.017 dex and -0.007 dex to the log(*gf*)s computed by M17 for the reference multiplets connected to the upper $4s4d\ ^3$D and the $4s5s\ ^3$S terms, respectively, based on our lifetime measurements on levels of these terms. This is essentially a correction to the radial matrix element for those reference multiplets. We adjusted the scale of all Ueda et al. (1982) log(*gf*)s by -0.030 dex, or -6.9%, to match, on average, the log(*gf*)s computed by M17 for the two reference multiplets with a small adjustment for our lifetime measurements. We then adjusted the M17 log(*gf*)s by +0.037 dex, +0.074 dex, and -0.014 dex for lines from the 4*s*5*d* $^3$D, 4*s*6*s* $^3$S, and 4$p^2$ $^3$P terms, respectively, to match on average, the Ueda et al. (1982) log(*gf*)s with the -0.030 dex rescaling. Note that the final correction of the M17 result for the 4$p^2$ $^3$P term, which has negligible IR residual, is small. This is an important confirmation of the accuracy of the Ueda et al. (1982) hook measurements and the M17 theoretical results for this strong multiplet. The single measurement by Smith (1988) of log(*gf*) = +0.292 for the 4302.53 Å line provides some additional confirmation.

Ueda et al. (1983) have lines in common with M17 for one additional UV multiplet connecting the upper 4*s*6*d* $^3$D term with lower 4*s*4*p* $^3$P term. For a reference, they used their earlier measurements on three lines of the blue multiplet studied by Ueda et al. (1982) connecting the upper 4$p^2$ $^3$P to lower 4*s*4*p* $^3$P term. For the $^3$P multiplet we use the M17 log(*gf*)s offset by -0.014 dex as specified above. We decided not to use the line at 3361.9124 Å, which had an inexplicably large discrepancy with M17 of 0.19 dex, and used the other three lines to determine an offset of -0.024 dex, or -5.4 %, for all Ueda et al. (1983) measurements. We adjusted the M17 log(*gf*)s of the UV multiplet connected to the upper 4*s*6*d* $^3$D term by +0.001 dex to match on average the Ueda et al. (1983) log(*gf*)s with the -0.024 dex rescaling. There are a variety of methods which could be used to renormalize the M17 log(*gf*)s using experimental results. We have chosen a method which preserves the excellent fine structure coupling of M17 and uses experimental lifetime measurements, and/or hook measurements to improve radial matrix elements. The rescaled results by Ueda et al. (1982, 1983) agree rather well with M17, and even the multiplet from the 4*s*6*s* $^3$S term near 3950 Å has offsets no worse than 0.07 dex, or 17%.

The absorption measurements by Smith & O'Neill (1975) include the same blue and red multiplets connected to the upper 4*s*4*d* $^3$D and 4*s*5*s* $^3$S terms used by Ueda et al. (1982) as reference multiplets. The above recommended log(*gf*)s for those reference multiplets are, on average, in agreement with Smith & O'Neill's log(*gf*)s if the 1975 results are offset of -0.02 dex.

The more recent measurements out of Oxford by Smith & Raggett (1981) and by Smith (1988) are both normalized with the line at 5349.47 Å with a log(*gf*) = -0.310 ± 0.020 dex, or ±4.7%. Relative *gf*s from absorption measurements were adjusted by the Oxford team before publication using a least-squares routine on closed loops of *gf*s. Relative uncertainties on transition probabilities of the strongest lines range down to 2.5%. Smith (1988) used the Hanle effect measurements by Hunter et al. (1985) and Hunter & Peck (1986) to set their absolute scale. We also checked their normalization using LIF radiative lifetimes from Havey et al. (1977) for the $4p^2\,^1S_0$ and $^1D_2$ levels, and found good agreement. The reader should note that Havey et al. (1977) used an old, incorrect swapped configuration assignment for the $4s6s\,^1S_0$ and $4p^2\,^1S_0$ levels. The uncertainties of the Havey et al. radiative lifetime measurements are significantly larger than the ~3% uncertainties on the Hanle Effect measurements by Hunter et al. (1985) and Hunter & Peck (1986) used by Smith (1988). Smith's normalization is used herein. In a few cases, where there is a discordance between log(*gf*)s from Smith & Raggett (1981) and from Smith (1988), we recommend the latter. Transition multiplets included in these two studies are indicated with green lines in Figure 1.

Table 8 has our recommended log(*gf*)s for 202 lines of Ca I ranging in wavelength from 2200 – 10,000 Å. Air wavelengths are given in the first column, upper and lower level energies are given in the second and third columns, respectively, and excitation potential (EP, lower level energy in eV) in the fourth column. The M17 theoretical log(*gf*) are given in the fifth column and the percent difference between the length and velocity forms of the calculations as reported by Yu & Derevianko (2018) appear in column six. Our recommended log(*gf*) are reported in the seventh column. Column eight gives the experimental source for lines where our recommended log(*gf*) differs from M17 and finally the ninth column gives the uncertainty for the experimentally augmented log(*gf*)s. In cases where there is a discordance between our recommended log(*gf*) and the M17 log(*gf*), the absolute difference is smaller than 0.2 dex except for six lines. Our conclusion is that the best modern theoretical transition probabilities for Ca I are nearly competitive with the best modern experiments. There are a great many, > 200, references reporting numerical data on transition probabilities of Ca I (e.g. the bibliography of the NIST ASD). Admittedly, our Table 8 only includes references from 1976 onward with two exceptions. This interval corresponds to the increasing use of tunable dye lasers to measure radiative lifetimes for normalization of transition probabilities. We also favored studies that covered many lines and continued for multiple years. Undoubtedly, there will be concerns that we should have included additional measurements in Table 8, but our goal was to compare the best modern theoretical transition probabilities with the best modern experimental transition probabilities.

## 5. CALCIUM ABUNDANCES IN THE SUN AND HD 84937

We used the recommended Ca I transition data in Table 8 to determine new Ca abundances in the solar photosphere and in the metal-poor main sequence star HD 84937. In general we followed the procedures of previous papers in this series of studies of Fe-group neutral and ionized species (e.g., Lawler et al. 2019, and references therein).

Ca I has a relatively simple electronic structure compared to many Fe-group species, yielding a relatively small number of strong absorption lines and other transitions that are very weak and undetectable in typical stellar spectra. As in previous papers of this series, we define relative line strength as

$$\mathrm{STR} \equiv \log(gf) - \theta\chi$$

where $\chi$ is the lower excitation energy of the transition and $\theta$ is the inverse temperature, 5040/T . The STR values computed here are applicable only to Ca I; attempts to combine them with Ca II would require at least an additional term to account for Saha neutral/ion ionization ratio. For this calculation we assume T = 5950 K, a compromise between the effective temperatures $T_{eff}$ of the Sun and HD 84937, the metal-poor main sequence turnoff star to be discussed in §5.2. The STR values are plotted versus wavelength in Figure 3. The general distribution of points in this plot is functionally similar to those seen in our previous neutral-species studies, e.g., V I (Lawler et al. 2014, Figure 3) and Co I (Lawler et al. 2015, Figure 3). But Ca is an alkaline earth element. Ca I has a simpler electronic energy structure than V I and Co I, leading to relatively few transitions, as can be seen in its strength plot.

### *5.1 Calcium in the Solar Photosphere*

In most papers of this series some effort has been made to identify all appropriate transitions for a species in the solar photosphere. This is not necessary here, because all Ca I lines that are useful for abundance analysis have been cataloged previously. In Figure 3 we draw a horizontal blue line to denote the approximate STR level for photospheric lines that have very small equivalent widths (*EW*s). Using reduced widths log(*RW*) ≡ log(*EW*/λ) which are nearly wavelength-independent, the line drawn at STR = –4.9 indicates the strength level for very weak lines on the linear part of the curve of growth, those with log(*RW*) ~ –6.0 (equivalent to 5 mÅ at 5000 Å). Transitions with smaller strengths are difficult to identify with certainty and are subject to larger abundance uncertainties. The solar photospheric spectral line compendium by Moore et al. (1966), covering the optical spectral region (2935–8770 Å) lists ~170 absorption features totally or partially attributable to Ca I. In Table 8, there are 87 lines with STR > –5.5 in the wavelength region 4000–8770 Å. Moore et al. identify 82 of these transitions in the solar spectrum, and all but one of the "missing" identifications are due to masking by very large transitions of other species. In the complex near-UV 3000–4000 Å region, 25 out of 30 lines with STR > –5.5 have solar identifications. Therefore a search for useful Ca I transitions in the solar spectrum is unnecessary; that task was done by Moore et al.

Most of the solar Ca I lines in the yellow-red spectral region (λ > 5000 Å) have no substantial contaminants and as such can be treated to single-line *EW* analyses; detailed synthetic spectrum computations are not necessary. However, the majority of these lines are strong enough, log(RW) ≥ –5.0, to be saturated and thus be on the "flat" part of the curve of growth. This means decreased sensitivity of their EWs to abundance, with increased sensitivity to microturbulent velocity $v_t$, and for the strongest lines some dependence on assumed damping parameters.

In Table 9 we list the *EW*s for the chosen Ca I lines. These were measured with SPECTRE[10] (Fitzpatrick & Sneden 1987), a specialized spectroscopic analysis code. The photospheric center-of-disk spectrum was the on-line BASS2000 version of Delbouille et al. (1973)[11]. We matched the observed lines with Gaussian and/or Voigt model profiles along with direct integrations to derive the *EW*s.

[10] http://www.as.utexas.edu/~chris/spectre.html

We derived Ca abundances from these EWs with the LTE line analysis code MOOG (Sneden 1973)[12]. The line parameters excitation energy and recommended log(*gf*) are given in Table 8. To ensure consistency with our previous studies of Fe-group and neutron-capture elements we employed the older Holweger & Müller (1974) model photospheric atmosphere in the computations. The derived abundances in log ε units[13] are listed in Table 9, and plotted versus line wavelength in the top panel of Figure 4. From EW measurement uncertainties the abundance uncertainties are typically ±0.03 for individual lines. Total uncertainties depend on chosen solar photospheric model atmospheres, adopted line analysis codes, and radiative transfer assumptions. These are not explored in our work, which concentrates on Ca I transition probabilities. For a good discussion of modeling issues see Scott et al. (2015). The mean elemental abundance from 39 Ca I transitions is <log ε> = 6.34 ±0.02 (σ = 0.09), as indicated in Figure 4. This value is in accord with other recent investigations of solar photospheric Ca abundance estimates: 6.34 ± 0.04 (Asplund et al. 2009), and 6.32 ± 0.03 (Scott et al. 2015)[14]. Recommended meteoritic abundances are slightly lower: log ε = 6.31 ± 0.02 (Lodders et al. 2009), log ε = 6.27 ± 0.03 (Lodders 2020).

We also enlisted Ca II to assist our solar Ca abundance determinations. Singly-ionized Ca is a light (only slightly relativistic) atomic ion with one valence electron outside of closed shells. The relativistic calculation including single, double, and triple excitations of Dirac-Fock wave functions by Safronova & Safronova (2011) yielded transition probabilities for lines of Ca II of high quality. Recently Kaur et al. (2021) tested the earlier work on Ca II by Safronova & Safronova and expanded the earlier work to include lines of Mg II, Sr II, and Ba II. Although the earlier work was motivated primarily by applications of Ca in atomic clocks, the excellent transition probabilities have important astrophysical applications. The transition probabilities by Kaur et al. (2021) agreed with those by Safronova & Safronova (2011) to well within 1% for nearly all lines.

---

[11] http://bass2000.obspm.fr/solar_spect.php

[12] http://www.as.utexas.edu/~chris/moog.html

[13] We use standard abundance notations. For elements X and Y, the relative abundances are written [X/Y] = $\log_{10}(N_X/N_Y)_{star} - \log_{10}(N_X/N_Y)_{\odot}$. For element X, the "absolute" abundance is written $\log_{10} \varepsilon(X) = \log_{10}(N_X/N_H) + 12$. Metallicity is defined as [Fe/H].

[14] This is their best value after consideration of NLTE and 3D effects. These authors also computed an LTE abundance with the Holweger & Müller (1974) solar model, finding log ε(Ca) = 6.34.

Ca II in the solar spectrum is dominated by Fraunhofer K (3933.7 Å) & H (3968.5 Å) and the near-IR triplet (8498.0, 8542.1, and 8662.2 Å) lines, but all of these features are extremely strong (log(RW) >> –4.0) and their line profiles are dominated by large damping wings. They are unreliable photospheric abundance indicators. However, Moore et al. (1966) identify nearly 20 other solar Ca II transitions. The ones with $\lambda > 4000$ Å arise from high excitation (EP $\geq$ 6.5 eV) lower excitation levels. Our laboratory investigation did not include Ca II, so we adopted transition probabilities from the theoretical computations of Safronova & Safronova (2011). The solar Ca II lines are generally weak and/or blended, so we derived Ca abundances from synthetic/observed spectrum matches. In panels a–c of Figure 5 we illustrate these matches for three of the lines. The original Delbouille et al. (1973) photospheric spectrum has a wavelength step size of 0.002 Å, so for plotting clarity we have shown points separated by 0.016 Å.

To construct the synthetic spectrum line lists we employed the *linemake* facility (Placco et al. 2021)[15], which begins with the Kurucz (2011, 2018)[16] atomic line database and substitutes/modifies/adds atomic transition data from the papers published in this series and related studies by the Wisconsin lab atomic physics group, and molecular transition data from the Old Dominion lab molecular physics group (e.g., Brooke et al. 2016, and references therein). These line lists were used to generate initial synthetic spectra to be compared to the observations. The observed/synthetic matches were generally reasonable, but we then adjusted the line wavelengths and transition probabilities subject to the following restrictions. If a line has been reported in a study co-authored by the Wisconsin group, including the Ca I transitions reported here, its line parameters were accepted without change. The lines without such laboratory information were adjusted in wavelength and log(*gf*) to produce best matches to the observed spectra. In this way the excellent overall matches seen in Figure 5, and in Figure 6 to be discussed in §5.2, demonstrate that spectrum features surrounding the Ca features of interest have good identifications and can be used to accurately assess their contamination in crowded spectral regions. These good overall fits are not intended to yield reliable abundances for any features except those of Ca.

---

[15] https://github.com/vmplacco/linemake
[16] http://kurucz.harvard.edu/linelists.html

In Table 10 we list the individual Ca II solar line abundances. From six lines we derive elemental means <log ε> = 6.34 ± 0.01 (σ = 0.02), consistent within the uncertainties to that derived by Scott et al. (2015). The very small line-to-line scatter is probably a reflection of just the synthetic/observed spectrum fitting, because all six of the lines arise from the same lower energy level: $3p^6 5p\ ^2P^o$. However, the agreement seen in Figure 4 between abundances derived for Ca I and Ca II is encouraging. Most of Ca exists in its ionized species, due to the low first ionization energy: IP(Ca I) = 6.11 eV. A simple Saha ionization balance in line-forming regions of the solar atmosphere (τ ~ 0.5) yields N(Ca II)/N(Ca I) ~ $10^3$. On the other hand, the excitation energy of the Ca II lines used for the solar analysis is high, EP = 7.51 eV, thus leading to significant temperature dependence of derived abundances. The excellent abundance agreement between neutral and ionized Ca species, combined with the similarity to previous photospheric and meteoritic results, suggests that the solar abundance derived with the new Ca I log(*gf*) values is reliable.

### *5.2 Calcium in HD 84937*

As in our previous papers on Fe-group transitions (e.g., Den Hartog et al. 2019, and references therein), to test application of the new transition probability data to low metallicity stars we chose the main sequence turnoff star HD 84937. The model parameters were those initially derived in Sneden et al. (2016): $T_{eff}$ = 6300 K, log g = 4.0, [Fe/H] = –2.15, and $v_t$ = 1.5 km $s^{-1}$. The optical (ESO VLT UVES, Dekker et al. 2000; McDonald 2.7m Tull Echelle, Tull et al. 1995) high-resolution spectra used here are discussed in detail in Sneden et al. The ultraviolet (UV) spectra were downloaded from the Mikulski Archive for Space Telescopes (MAST) and processed automatically by the CALSTIS pipeline. These spectra were obtained using the Space Telescope Imaging Spectrograph (STIS, Kimble et al. 1998; Woodgate et al. 1998) on board the Hubble Space Telescope, as part of program GO-14161 (PI Peterson). They cover 1879–3143 Å at a spectral resolving power of 114,000. Our abundance computations use the MOOG version that includes scattering in the continuum source functions (Sobeck et al. 2011), but in reality the scattering terms are not large even in the *UV* spectral domain of the warm, high gravity HD 84937; $H^-$ remains the dominant continuum opacity source, followed by H I Balmer continuum opacity blueward of the Balmer jump (λ < 3646 Å).

Identification of useful Ca I lines was as straightforward as in the solar case described in §5.1. Nearly all of the transitions in Table 8 are in the optical domain, λ > 3250 Å, and all of the strong lines have solar identifications. The only UV transitions in this study are nine arising from the Ca I ground state.

We measured *EW*s and derived abundances as described in §5.1. These values are listed in Table 9 and plotted as a function of wavelength in the bottom panel of Figure 4. Additionally, we used synthetic spectrum computations to extract abundances from the ground-state lines in the *UV*. These transitions are important because most of the electrons in Ca I exist in the ground state instead of the excited states (whose minimum EP is 1.69 eV). There are three ground-state lines strong enough for detection in HD 84937: 2200.73, 2275.47, and 2398.56 Å. These transitions are indicated by blue lines in Figure 1 connecting to the 4*s*6*p*, 4*snp* and 4*s*7*p* $^1P_1$ levels. Unfortunately, the 2200.73 Å transition is completely masked by the very strong Fe I 2200.72 Å line, aided by a strong Ni I line at 2200.68 Å. In Figure 6 we show observed and synthetic spectra for the other two ground-state lines. The 2275.47 Å line has only a very weak contaminant: Fe II 2275.47 Å, which gives rise to the ~2% absorption seen at this wavelength in the no-Ca synthesis. The Fe II transition is real, as it is listed in the NIST ASD (Kramida et al. 2019**;** the line is from an unpublished line list accompanying the lab study by Nave & Johansson 2013). However, no lab transition probability has been published, and the log(*gf*) value adopted for this line comes from the Kurucz database, one of the Fe II lines with "semiempirical" computed values (e.g., Kurucz 1988, and references therein). The contamination of the Ca I line caused by this Fe II line is probably very small, but it remains somewhat uncertain. Finally, the 2398.56 Å transition, displayed in the bottom panel of Figure 6, is somewhat blended and must be analyzed via synthetic spectra. The HD 84937 mean abundance from the 49 Ca I lines is $\langle \log \varepsilon \rangle = 4.48 \pm 0.01$ ($\sigma = 0.05$). The two UV ground-state lines, with $\langle \log \varepsilon \rangle = 4.44$ ($\sigma = 0.03$ from the synthetic/observed spectrum fits), are in reasonable agreement with the overall average.

We derived abundance sensitivities to model atmosphere uncertainties in the manner computed by Sneden et al. (2016). In that paper, consideration of the many literature studies of HD 84937 suggested that parameter uncertainties are $\sigma(T_{eff}) \approx \pm80$ K, $\sigma(\log g) \approx \pm0.15$, $\sigma([M/H]) \approx \pm0.15$, and $\sigma(v_t) \approx \pm0.1$ km s$^{-1}$. We altered the adopted HD 84937 model to assess the Ca I abundance changes in response to these parameter shifts. For temperature changes $\Delta(T_{eff}) = \pm150$ K, $\Delta(\log \varepsilon(\text{Ca I})) \approx \pm0.07$; for surface gravity changes $\Delta(\log g) = \pm0.25$, $\Delta(\log \varepsilon(\text{Ca I})) \approx \mp0.01$; for metallicity changes $\Delta([M/H]) = \pm0.25$, $\Delta(\log \varepsilon(\text{Ca I})) < 0.01$; and for microturbulent velocity changes $\Delta(v_t) = \pm0.15$ km s$^{-1}$, $\Delta(\log \varepsilon(\text{Ca I})) \approx \mp0.01$. Clearly the greatest abundance sensitivity is to $T_{eff}$. However, when these log ε are converted to [Ca/Fe] abundance ratios (§6), the similar sensitivities of Fe I to $T_{eff}$ uncertainties yields relatively small dependence of Ca I abundances on model parameters, as argued in Sneden et al. and other literature sources.

Many Ca II lines are accessible in HD 84937. We show three examples of synthetic/observed spectrum matches in panels d–f of Figure 5. Although analyses of these lines are relatively straightforward, the spectrum of the 2132.30 Å feature provides an illustration of the uncertainties of our procedure. The Ca II line is primarily blended with Al I 2132.39 Å. This contaminant is cataloged in the NIST ASD (from Penkin & Shabanova 1965) but without a transition probability, so our syntheses relied on a best empirical fit to set the Al I strength. Additionally, at about 2132.65 Å there is a missing transition in our synthetic spectrum line list. It does not have a plausible match in the Kurucz (2011, 2018) line compendium. The absence of a good match here serves as a cautionary note to our analyses: we cannot account for contaminants to the lines of interest if we have no empirical knowledge of their existence. With that caution in mind, it is clear that the 12 Ca II lines yield internal agreement, leading to $\langle \log \varepsilon \rangle = 4.51 \pm 0.02$ ($\sigma$ = 0.07). This value agrees within mutual uncertainties with the mean abundance from Ca I.

Finally, repeating for Ca II the parameter sensitivity computations, for temperature changes $\Delta(T_{eff}) = \pm150$ K, $\Delta(\log \varepsilon(\text{Ca II})) \approx \pm0.06$; for surface gravity changes $\Delta(\log g) = \pm0.25$, $\Delta(\log \varepsilon(\text{Ca II})) \approx \mp0.03$; for metallicity changes $\Delta([M/H]) = \pm0.25$, $\Delta(\log \varepsilon(\text{Ca II})) \approx \mp0.02$; and for microturbulent velocity changes $\Delta(v_t) = \pm0.15$ km s$^{-1}$, $\Delta(\log \varepsilon(\text{Ca II})) \approx \mp0.06$. The abundance shift with $T_{eff}$ change is largely driven by the sensitivity of $H^-$ continuous opacity to temperature in this HR diagram domain, which affects also the Ca I abundance as discussed above. The relatively large sensitivity of log ε(Ca II) to microturbulence is due to the large line strengths of the Ca II lines used here.

*5.3 Discussion*

Utilizing our new Ca abundance determination and our previously derived Fe abundance for the well-studied halo star HD 84937 (Sneden et al. 2016), we derive a value of [Ca/Fe] = +0.46 for this star. Based upon the discussion in the previous section regarding uncertainties in atmospheric parameters, we adopt a total error bar of ±0.07 in the log ε abundance of Ca. This error estimate does not include any possible NLTE effects for Ca or for Fe, but this may be an overestimate of the uncertainty in the ratio of [Ca/Fe] (see Sneden et al. 2016). We also look forward to determining this abundance ratio in additional stars with the same level of precision, utilizing the new experimentally determined atomic physics parameters. We compare this newly obtained value in Figure 7 with other low-metallicity halo stars from the survey of Roederer et al. (2014). That survey employed similar stellar parameters and model assumptions as our current calculations. It is clear from the figure that the abundance value for HD 84937 is entirely consistent with stars of similar metallicity in this survey. We have also made a similar abundance comparison with other data surveys that derived [Ca/Fe] (i.e., Cayrel et al. 2004, Barklem et al. 2005, Cohen et al. 2008, Lai et al. 2008 and Yong et al. 2013) employing different stellar parameters than that of Roederer et al. (2014). Again, we find the newly derived abundance value of [Ca/Fe] for HD 84937 to be generally consistent with the measured values of stars of similar metallicity in these other stellar data surveys.

These types of abundance comparisons that illuminate Galactic chemical evolution serve as probes of star formation and nucleosynthesis over the history of the Galaxy (e.g., Kobayashi et al. 2020). For example, the abundance data shown in Figure 7 indicate that the [Ca/Fe] abundance ratios are relatively flat over most of the metallicity range studied, with a standard deviation in [Ca/Fe] of approximately ±0.08 for $-3 < [Fe/H] < -2$, which is within the range of measurement uncertainty. This consistency indicates a steady (and perhaps related) production mechanism for both Ca and Fe. Ca is produced in explosive burning in higher mass stars, while Fe production comes from core collapse supernovae early in the history of the Galaxy (e.g., Thielemann et al. 1996).

We note in the figure that at the lowest values of [Fe/H], there appears to be a slight rising trend in [Ca/Fe] with a fair amount of scatter, which may be related to differences in astrophysical or nucleosynthetic outcomes. Confirming whether and to what extent this trend is occurring is significant to understanding early Galactic nucleosynthesis and the (apparent) concomitant decrease in the formation of iron. It suggests that perhaps at some point the Ca and Fe production might be decoupled, possibly as a result of different stellar mass ranges or environments for the production of these two elements. Further, significant scatter at those very low metallicities might also suggest varying mass ranges for the First Stars and/or inhomogeneity in early Galactic star forming regions.

We also note that Type Ia supernovae contribute greatly to Fe production at higher metallicities and more recent epochs. The resulting downward trend of [Ca/Fe] as metallicities become larger (starting near [Fe/H] = -1) is evident in the figure. Clearly additional studies of Ca in the very lowest metallicity stars (less than -4), where there are little data, are warranted and would provide insight into the earliest production of these elements to address the question of whether, for example, Ca and Fe production differ greatly or are unrelated in the earliest, and presumably massive, stars.

We are endeavoring to obtain new Ca (and other elemental) abundance determinations for a range of metal-poor stars to help probe that early Galactic history (A. M. Boesgaard et al., in preparation). It will also be important to compare the Ca data with other $\alpha$ element abundances, such as Si and Mg, to understand better both the nature of element production and star formation history at the earliest Galactic times.

## 6. SUMMARY

The goals of this study have been to assemble the best experimental data – both new to this study and previously published – to test the reliability of modern theoretical transition probabilities for the light alkaline earth element Ca, and to assemble the most accurate and comprehensive line list of Ca I transition probabilities as is currently possible. We present new radiative lifetimes and BF (or BR) measurements for transitions of Ca I. These, combined with other quality experimental data are used to test the ab initio CI+MBPT calculations of M17. We find that for most transitions the calculations stand up well to the comparison, with the M17 log(*gf*)s falling within 0.2 dex of the experimental log(*gf*)s for all but six lines. We use a combination of the theoretical results augmented with experiment to produce a set of recommended log(*gf*)s for 202 transitions of Ca I ranging in wavelength from 2200 – 10,000 Å . These data have been applied to the determination of the abundances of Ca in the Sun and the metal-poor star HD 84937.

## ACKNOWLEDGEMENTS

This work is supported by NSF grants AST-1814512 (E.D.H. and J.E.L), AST-1616040 (C.S.), AST-1815403 (I.U.R.), PHY 14-30152 (Physics Frontier Center/JINA-CEE); and NASA grants NNX16AE96G (J.E.L.), HST-GO-14765, and HST-GO-15657 (J.E.L. and I.U.R.). EDH and JEL would like to acknowledge the participation of Alex Scherer in the early stages of the radiative lifetime measurements.

Facility: HST (STIS), VLT (UVES)

Software: linemake (Placco et al. 2021), MOOG (Sneden 1973), SPECTRE (Fitzpatrick & Sneden 1987).

# FIGURE CAPTIONS

Figure 1. Partial Grotrian diagram of Ca I showing all but two high-lying configurations considered by M17. Configuration (and term where applicable to entire column) of each vertical set of levels is given along the bottom axis. The singlet/triplet structure is illustrated. The diagram is to-scale in the vertical direction except for the spacing of the levels in each triplet, which is exaggerated for the purpose of illustration. Also shown are lines connecting terms which indicate the transition multiplets from experimental studies discussed in the text. Legend abbreviations: UW – this study; Ueda 82,83 – Ueda et al. (1982) and Ueda et al. (1983); S&R81 – Smith & Raggett (1981) and Sm88 – Smith (1988); Park 76 – Parkinson et al. (1976); Vogt 07 – Vogt et al. (2007).

Figure 2. Logarithmic comparison of measured (UW) and LS theory branching fractions (BFs) with those of M17. The x symbols indicate log(BF) differences between this study (UW) and M17, and open circles indicate those differences between LS theory and M17. The solid horizontal line corresponds to perfect agreement with M17. Panels a-e present the data of Tables 2-6, respectively. All panels are displayed with the same vertical scale to emphasize that neutral Ca obeys LS coupling to a high degree (panels a-d), and where there is a divergence from LS theory due to mixing of levels (panel e), M17 has calculated that mixing accurately.

Figure 3. Relative strengths STR for the Ca I lines of this study. All lines with STR > –9 are shown with black dots, and those used in the solar abundance analysis are also circled in red. The 14 lines with STR < –9 not shown in this figure are too weak to be of interest in astronomical applications. The 3000 Å vertical line denotes the atmospheric wavelength cutoff for ground-based spectroscopy, and the horizontal line at STR ≈ –4.9 denotes the strengths of very weak solar lines, log(RW) ~ –6 (EW = 5 mÅ at λ = 5000 Å).

Figure 4. Line abundances for Ca I and Ca II in the solar photosphere (top panel) and HD 84937 (bottom panel). The species are distinguished by blue colors for Ca I and red for Ca II. In each panel horizontal lines are drawn to indicate the abundance mean and standard deviation σ derived for Ca I.

Figure 5. Observed and synthetic spectra of Ca II lines in the solar photosphere (panels a–c) and HD 84937 (panels d–f). In each panel the observations are indicated with points and the synthetic spectra with lines colored to represent different logarithmic abundances of Ca. The black line in each panel is for the abundance that best matches the synthetic and observed spectrum for that Ca II feature. Colors blue, green, and orange represent the best abundance offset by –0.6, –0.3, and +0.3 dex, respectively. The red line shows the effect of eliminating the Ca II feature completely.

Figure 6. Observed and synthetic spectra of two ground-state (EP = 0.0 eV) transitions of Ca I in the UV spectrum of HD 84937. The symbols and colors are as in Figure 5.

Figure 7. Calcium abundances [Ca/Fe] plotted versus metallicity [Fe/H]. For plotting clarity we have chosen to compare our HD 84937 abundance to the general trend from one survey, Roederer et al. (2014).

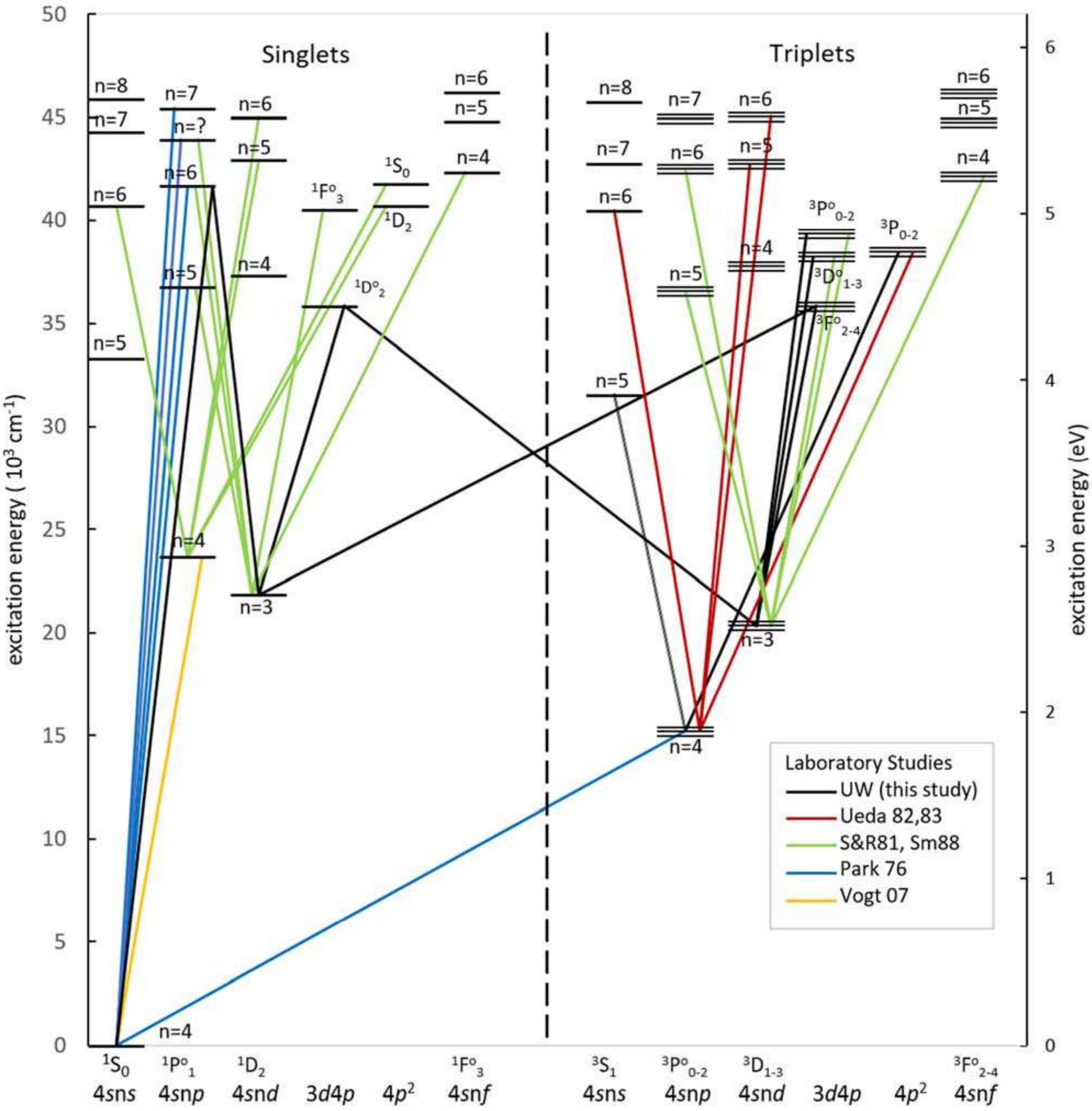

Figure 1. Partial Grotrian diagram of Ca I showing all but two high-lying configurations considered by M17. Configuration (and term where applicable to entire column) of each vertical set of levels is given along the bottom axis. The singlet/triplet structure is illustrated. The diagram is to-scale in the vertical direction except for the spacing of the levels in each triplet, which is exaggerated for the purpose of illustration. Also shown are lines connecting terms which indicate the transition multiplets from experimental studies discussed in the text. Legend abbreviations: UW – this study; Ueda 82,83 – Ueda et al. (1982) and Ueda et al. (1983); S&R81 – Smith & Raggett (1981) and Sm88 – Smith (1988); Park 76 – Parkinson et al. (1976); Vogt 07 – Vogt et al. (2007).

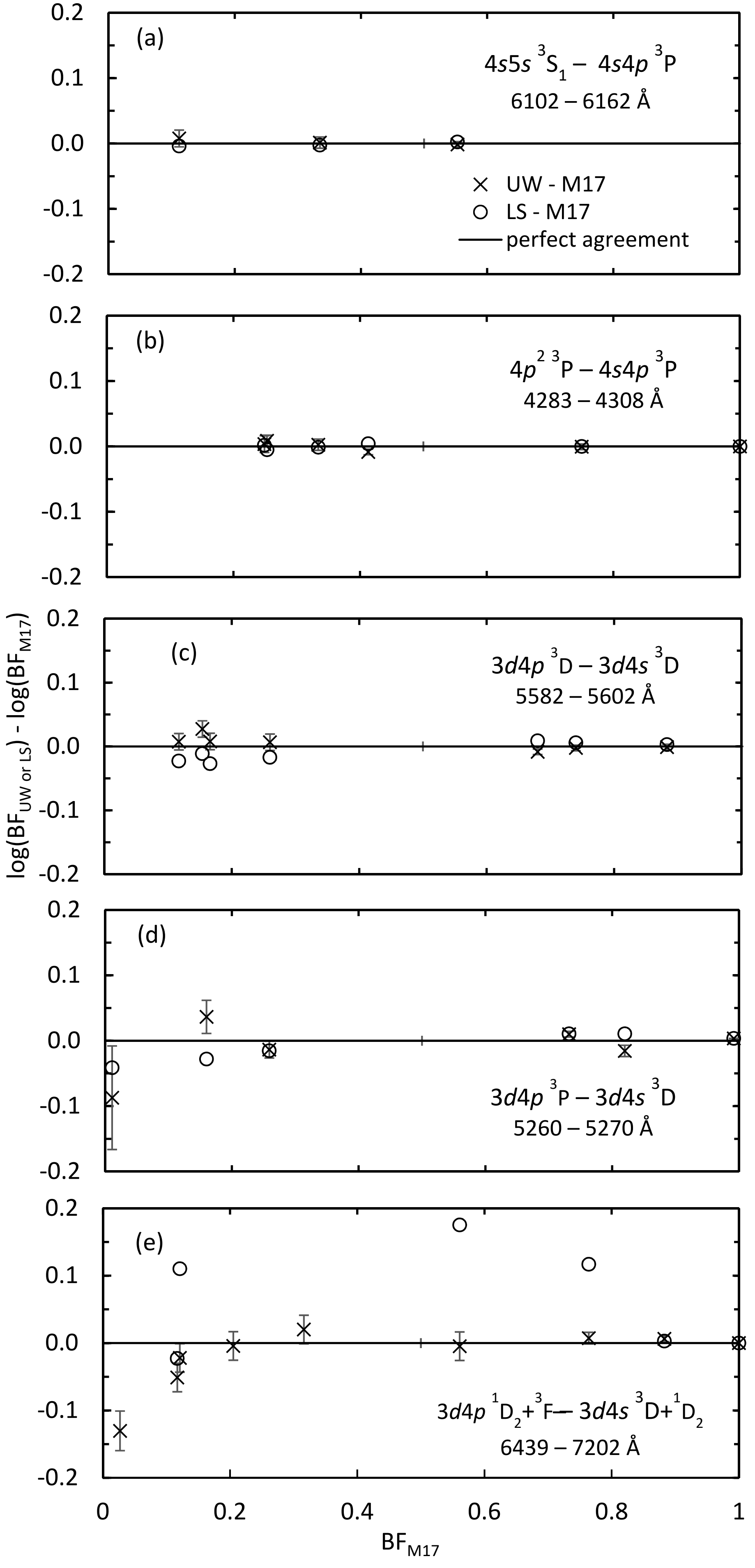

Figure 2. Logarithmic comparison of measured (UW) and LS theory branching fractions (BFs) with those of M17. The x symbols indicate log(BF) differences between this study (UW) and M17 and open circles indicate differences between LS theory and M17. The solid horizontal line corresponds to perfect agreement with M17. Panels a-e present the data of Tables 2-6, respectively. All panels are displayed with the same vertical scale to emphasize that neutral Ca obeys LS coupling to a high degree (panels a-d), and where there is a divergence from LS theory due to mixing of levels (panel e), M17 has calculated that mixing accurately.

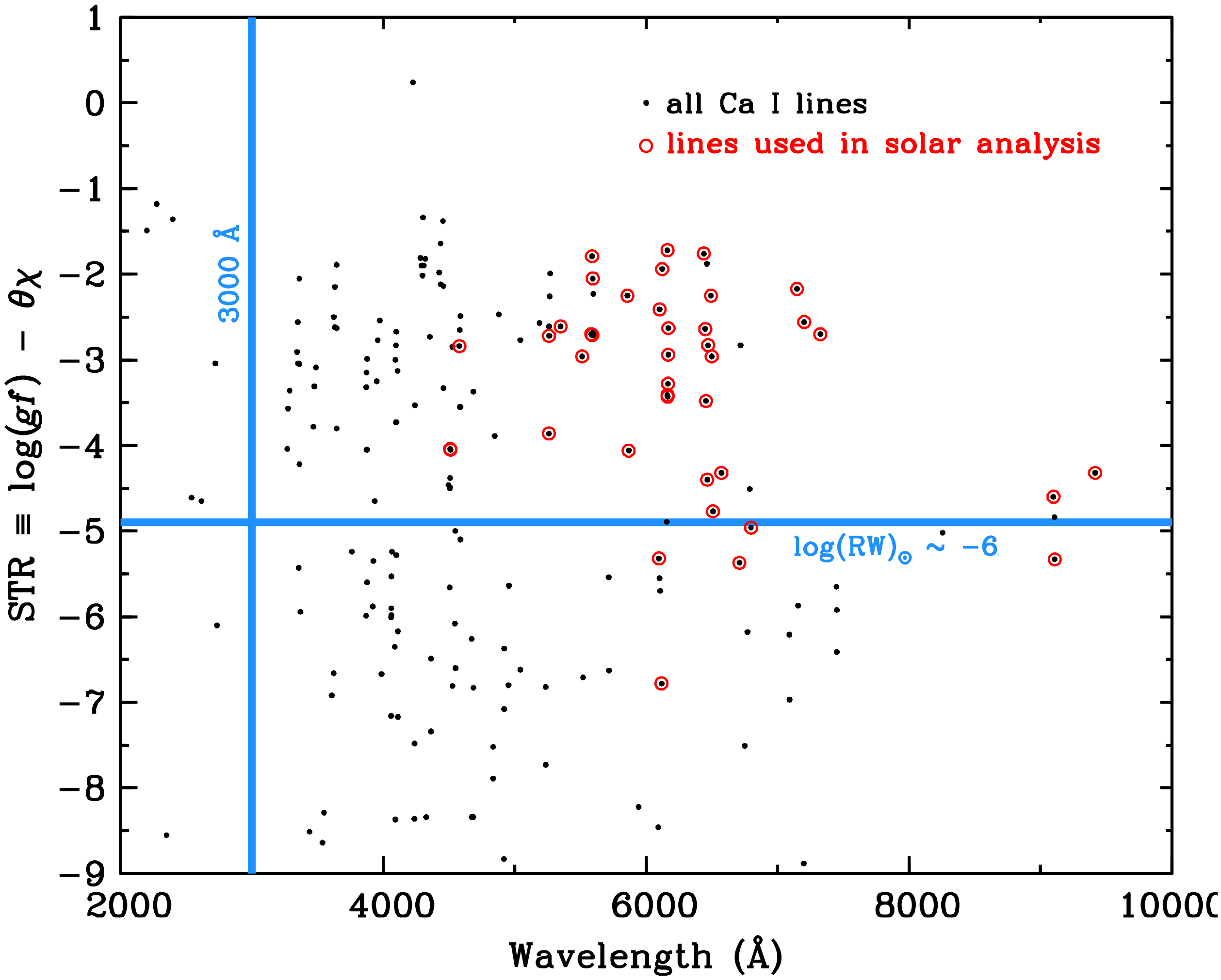

Figure 3. Relative strengths STR for the Ca I lines of this study. All lines with STR > –9 are shown with black dots, and those used in the solar abundance analysis are also circled in red. The 14 lines with STR < –9 not shown in this figure are too weak to be of interest in astronomical applications. The 3000 Å vertical line denotes the atmospheric wavelength cutoff for ground-based spectroscopy, and the horizontal line at STR ≈ –4.9 denotes the strengths of very weak solar lines, log(RW) ~ –6 (EW = 5 mÅ at λ = 5000 Å).

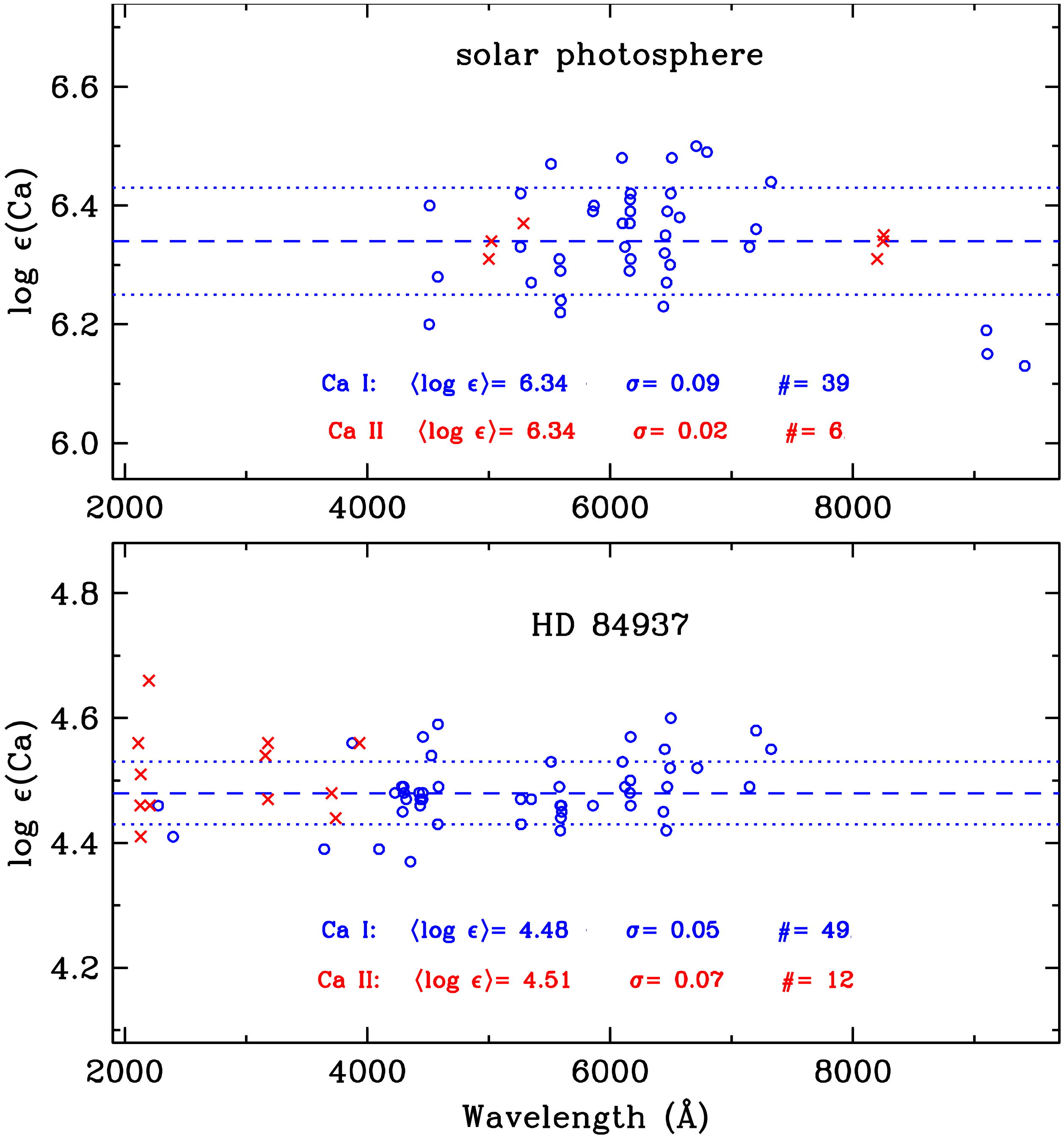

Figure 4. Line abundances for Ca I and Ca II in the solar photosphere (top panel) and HD 84937 (bottom panel). The species are distinguished by blue colors for Ca I and red for Ca II. In each panel horizontal lines are drawn to indicate the abundance mean and standard deviation σ derived for Ca I.

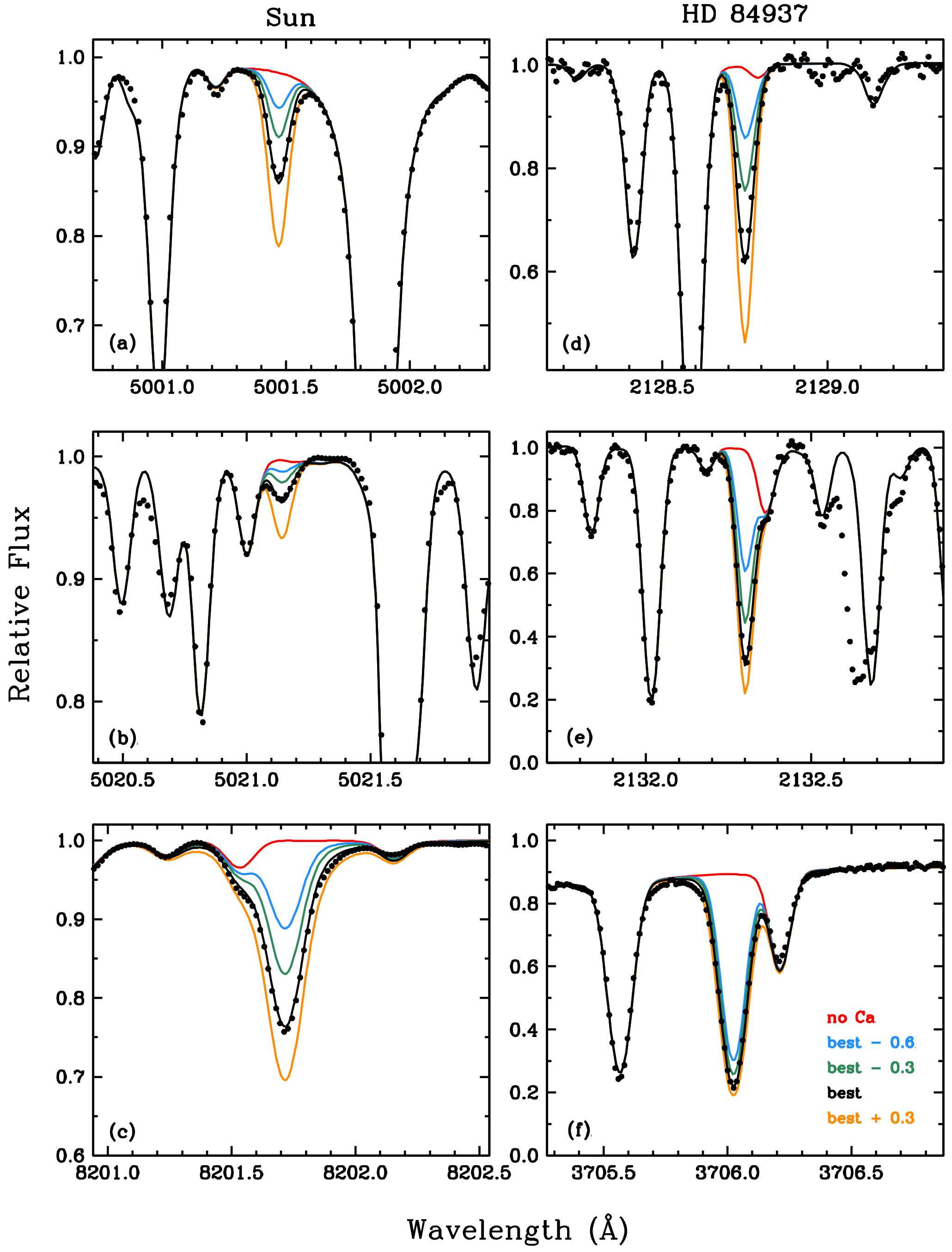

Figure 5. Observed and synthetic spectra of Ca II lines in the solar photosphere (panels a–c) and HD 84937 (panels d–f). In each panel the observations are indicated with points and the synthetic spectra with lines colored to represent different logarithmic abundances of Ca. The black line in each panel is for the abundance that best matches the synthetic and observed spectrum for that Ca II feature. Colors blue, green, and orange represent the best abundance offset by –0.6, –0.3, and +0.3 dex, respectively. The red line shows the effect of eliminating the Ca II feature completely.

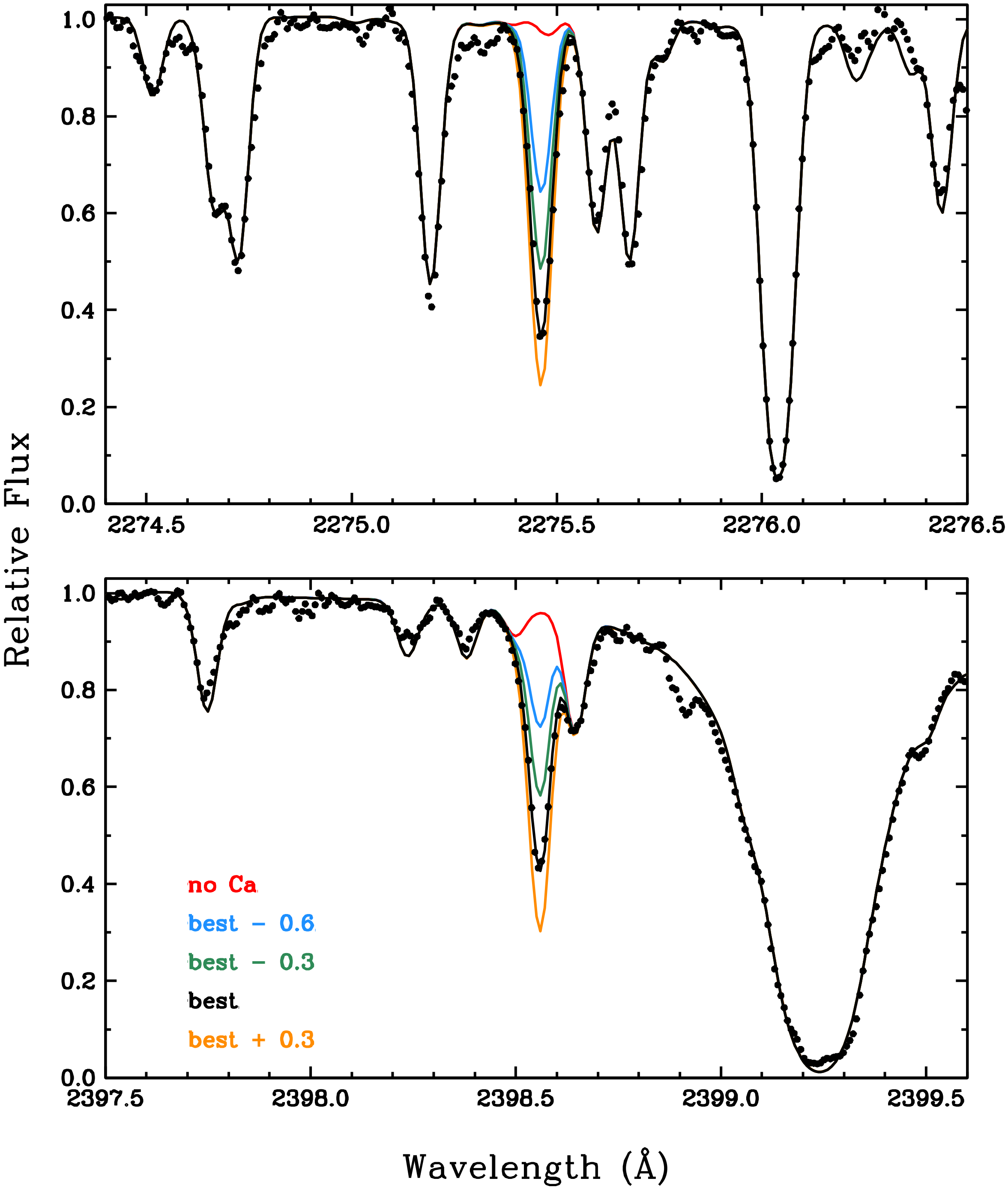

Figure 6. Observed and synthetic spectra of two ground-state (EP = 0.0 eV) transitions of Ca I in the UV spectrum of HD 84937. The symbols and colors are as in Figure 5.

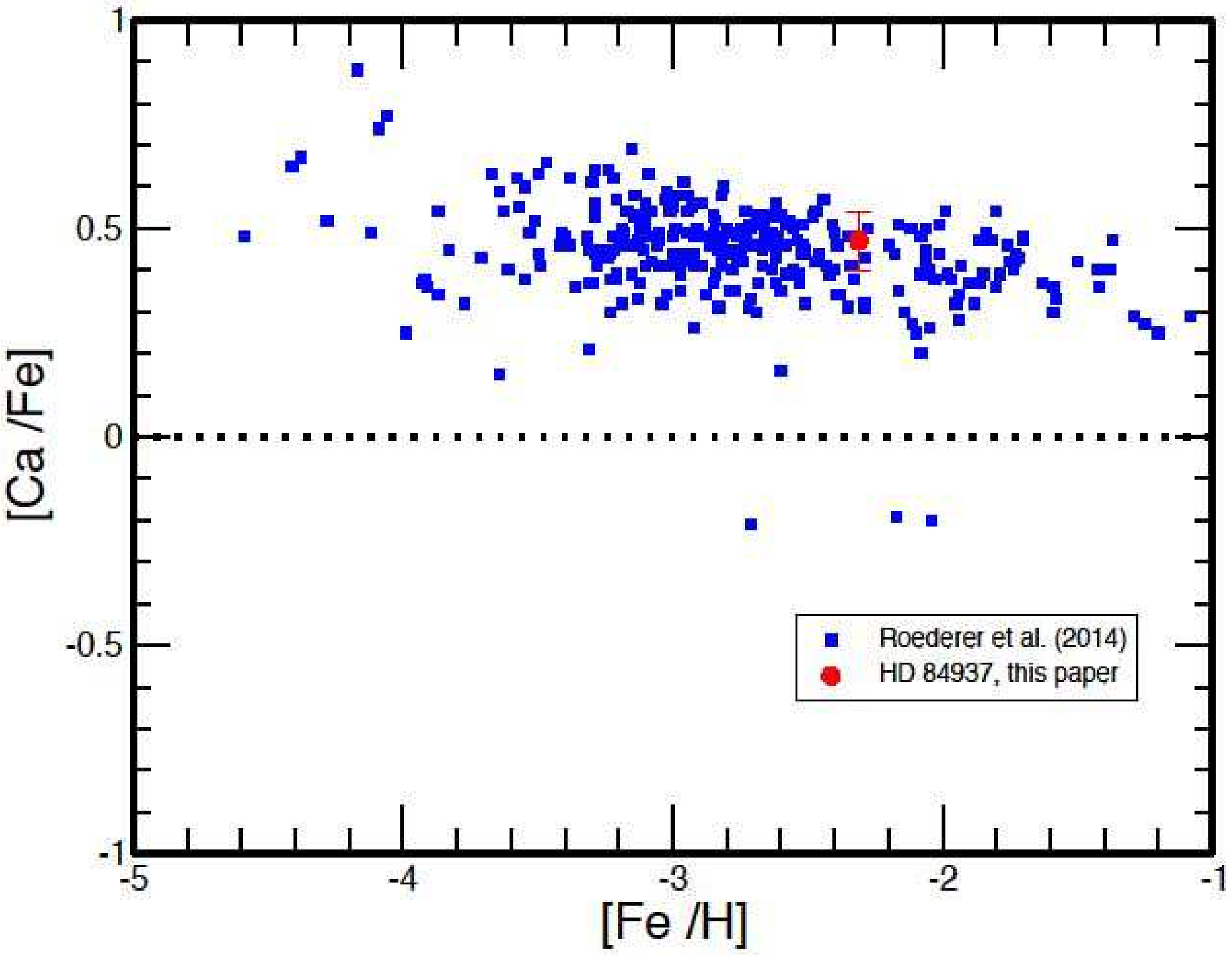

Figure 7. Calcium abundances [Ca/Fe] plotted versus metallicity [Fe/H]. For plotting clarity we have chosen to compare our HD 84937 abundance to the general trend from one survey, Roederer et al. (2014).

This page intentionally left blank.

Table 1
Radiative lifetimes for levels of neutral Ca.

| Configuration[a] | Term[a] | Level Energy[a] ($cm^{-1}$) | Radiative lifetimes (ns) UW[b] | M17 theory | Havey et al. (1977) | Hansen (1983) | Aldenius et al. (2009) | Jönsson (1984) | Other | Other Ref. |
|---|---|---|---|---|---|---|---|---|---|---|
| $3p^6 4s4p$ | $^1P^o_1$ | 23652.304 | 4.64 ± 0.23 | 4.6 | 4.7 ± 0.5 | 4.6 ± 0.2 | … | … | 4.6386 ± 0.0017 | 1 |
| | | | | | | | | | 4.65 ± 0.04 | 2 |
| | | | | | | | | | 4.49 ± 0.07 | 3 |
| | | | | | | | | | 4.6 ± 0.2 | 4 |
| $3p^6 4s5s$ | $^3S_1$ | 31539.495 | 12.3 ± 0.6 | 12.1 | 11.7 ± 0.6 | … | 13.8 ± 1.2 | … | 10.7 ± 1.0 | 5 |
| | | | | | | | | | 12.4 ± 0.5 | 6 |
| $3p^6 4s4d$ | $^1D_2$ | 37298.287 | … | 62.3 | 80.2 ± 9.6 | … | … | … | 63 ± 10 | 7 |
| $3p^6 4s4d$ | $^3D_1$ | 37748.197 | 12.2 ± 0.6 | 12.7 | … | … | 13.6 ± 1.0 | 13.0 ± 0.5 | 12.3 ± 1.0 | 5 |
| | $^3D_2$ | 37751.867 | 12.2 ± 0.6 | 12.8 | … | … | 14.1 ± 1.1 | 13.0 ± 1.5 | … | … |
| | $^3D_3$ | 37757.449 | 12.3 ± 0.6 | 12.8 | … | … | 15.4 ± 1.2 | 13.0 ± 2.5 | … | … |
| $3p^6 4p^2$ | $^3P_0$ | 38417.543 | 5.80 ± 0.29 | 5.5 | … | … | … | 6.9 ± 0.4 | … | … |
| | $^3P_1$ | 38464.808 | 5.63 ± 0.28 | 5.5 | … | … | … | 6.9 ± 0.4 | … | … |
| | $^3P_2$ | 38551.558 | 5.66 ± 0.28 | 5.5 | … | … | … | 6.9 ± 0.4 | … | … |
| $3p^6 4s6s$ | $^3S_1$ | 40474.241 | 31.7 ± 1.6 | 30.6 | … | … | … | … | … | … |
| $3p^6 3d4p$ | $^1F^o_3$ | 40537.893 | … | 137.3 | … | … | … | … | 58.3 ± 2 | 8 |
| $3p^6 4s6s$ | $^1S_0$ | 40690.435 | … | 142.0 | 88.9 ± 3.0 | … | … | … | … | … |
| $3p^6 4p^2$ | $^1D_2$ | 40719.847 | … | 16.9 | 14.9 ± 0.9 | … | … | 16.3 ± 2.0 | 14.47 ± 0.14 | 7 |
| $3p^6 4s6p$ | $^1P^o_1$ | 41679.008 | 20.6 ± 1.0 | 27.7 | | … | … | … | … | … |
| $3p^6 4p^2$ | $^1S_0$ | 41786.276 | … | 12.6 | 12.6 ± 0.8 | … | … | 14.4 ± 0.7 | … | … |
| $3p^6 4s4f$ | $^1F^o_3$ | 42343.587 | … | 26.4 | … | … | … | … | 24.6 ± 1.4 | 9 |
| $3p^6 4s5d$ | $^3D_1$ | 42743.002 | 24.8 ± 1.2 | 24.7 | … | … | 27.0 ± 2.0 | … | … | … |
| | $^3D_2$ | 42744.716 | 24.9 ± 1.2 | 24.7 | … | … | 27.0 ± 2.0 | … | … | … |
| | $^3D_3$ | 42747.387 | 24.9 ± 1.2 | 24.8 | … | … | 27.0 ± 2.0 | … | … | … |
| $3p^6 4s5d$ | $^1D_2$ | 42919.053 | … | 19.6 | 22.2 ± 1.3 | … | … | … | 21.3 ± 0.7 | 7 |
| $3p^6 4snp$ | $w^1P^o_1$ | 43933.477 | 15.7 ± 0.8 | 13.2 | … | … | … | … | … | … |
| $3p^6 4s7s$ | $^1S_0$ | 44276.538 | … | 92.5 | 62.4 ± 4.3 | 63.4 ± 2.9 | … | … | … | … |
| $3p^6 4s5f$ | $^1F^o_3$ | 44804.878 | … | 43.3 | … | … | … | 60 ± 3 | 62.1 ± 2 | 9 |
| $3p^6 4s6d$ | $^1D_2$ | 44989.830 | … | 91.7 | … | 58.9 ± 2.5 | … | … | … | … |
| $3p^6 4s7p$ | $^1P^o_1$ | 45425.358 | 43.3 ± 2.2 | 43.7 | … | … | … | … | … | … |
| $3p^6 4s8s$ | $^1S_0$ | 45887.200 | ... | 156.7 | 118.3 ± 6.3 | 124.9 ± 6.0 | ... | ... | ... | ... |
| $3p^6 4s6f$ | $^1F^o_3$ | 46182.399 | ... | 96.7 | ... | ... | ... | ... | 99 ± 4 | 9 |
| $3p^6 4s7d$ | $^1D_2$ | 46200.13 | ... | ... | ... | 224 ± 12 | ... | ... | ... | ... |
| $3p^6 4s9s$ | $^1S_0$ | 46835.055 | ... | ... | ... | 206 ± 11 | ... | ... | ... | ... |
| $3p^6 4s10s$ | $^1S_0$ | 47437.471 | ... | ... | ... | 294 ± 17 | ... | ... | ... | ... |
| $3p^6 3d5s$ | $^1D_2$ | 47449.083 | ... | ... | ... | 81.2 ± 6.9 | ... | ... | ... | ... |
| $3p^6 4s9d$ | $^1D_2$ | 47812.39 | ... | ... | ... | 232 ± 12 | ... | ... | ... | ... |
| $3p^6 4s11s$ | $^1S_0$ | 47843.76 | ... | ... | ... | 420 ± 29 | ... | ... | ... | ... |
| $3p^6 4s10d$ | $^1D_2$ | 48083.41 | ... | ... | ... | 271 ± 16 | ... | ... | ... | ... |
| $3p^6 4s12s$ | $^1S_0$ | 48130.75 | ... | ... | ... | 594 ± 46 | ... | ... | ... | ... |
| $3p^6 4s11d$ | $^1D_2$ | 48290.85 | ... | ... | ... | 425 ± 36 | ... | ... | ... | ... |
| $3p^6 4s13s$ | $^1S_0$ | 48340.75 | ... | ... | ... | 808 ± 72 | ... | ... | ... | ... |
| $3p^6 4s14s$ | $^1S_0$ | 48499.14 | ... | ... | ... | 720 ± 65 | ... | ... | ... | ... |
| $3p^6 3d^2$ | $^3P_2$ | 48563.522 | 8.7 ± 0.4 | ... | ... | ... | ... | ... | ... | ... |

**Notes:**

[a]configuration, term and level energy from NIST ASD (Kramida, et al. 2019)

[b]The uncertainty on all UW lifetimes (present study) is ±5%

Other References: (1) Vogt et al. (2007) (2) Degenhardt et al. (2003) (3) Kelly & Mathur (1980) (4) Hannaford & Lowe (1981) (5) Gornik et al. (1973) (6 )Major et al. (1985) (7) Hunter et al. (1985) (8) Hunter & Peck (1986) (9) Bhatia (1986)

Table 2

Emission BF measurements on a multiplet from the upper 4*s*5*s* $^3S_1$ level to the 4*s*4*p* $^3P$ term.

| $E_{upp}$ ($cm^{-1}$) | $J_{upp}$ | $E_{low}$ ($cm^{-1}$) | $J_{low}$ | $\lambda_{air}$ (Å) | BF this expt. | Unc. (%) | BF A09[a] | Unc. (%) | BF M17 | BF pure LS[b] |
|---|---|---|---|---|---|---|---|---|---|---|
| 31539.495 | 1 | 15157.901 | 0 | 6102.7227 | 0.114 | 3 | 0.115 | 3 | 0.112 | 0.111 |
| 31539.495 | 1 | 15210.063 | 1 | 6122.2172 | 0.336 | 2 | 0.339 | 2 | 0.335 | 0.333 |
| 31539.495 | 1 | 15315.943 | 2 | 6162.1730 | 0.551 | 1 | 0.546 | 2 | 0.553 | 0.556 |

Notes

[a] A09: Aldenius et al. (2009)

[b] LS BFs are calculated from perturbation theory.  See, for example, Appendix I of Cowan (1981) for a tabulation of LS line strengths.

Table 3

Emission BF measurements on a multiplet from the upper $4p^2$ $^3$P term to the $4s4p$ $^3$P term.

| $E_{upp}$ (cm$^{-1}$) | $J_{upp}$ | $E_{low}$ (cm$^{-1}$) | $J_{low}$ | $\lambda_{air}$ (Å) | BF This expt. | Unc. (%) | BF M17 | BF pure LS |
|---|---|---|---|---|---|---|---|---|
| 38417.543 | 0 | 15210.063 | 1 | 4307.7439 | 1.000 | 0 | 1.000 | 1.000 |
| 38464.808 | 1 | 15157.901 | 0 | 4289.3668 | 0.336 | 2 | 0.334 | 0.333 |
| 38464.808 | 1 | 15210.063 | 1 | 4298.9883 | 0.258 | 2 | 0.253 | 0.250 |
| 38464.808 | 1 | 15315.943 | 2 | 4318.6517 | 0.405 | 1 | 0.413 | 0.417 |
| 38551.558 | 2 | 15210.063 | 1 | 4283.0106 | 0.251 | 3 | 0.249 | 0.250 |
| 38551.558 | 2 | 15315.943 | 2 | 4302.5278 | 0.749 | 1 | 0.750 | 0.750 |

Table 4

Emission BF measurements on the multiplet from the upper $3d4p$ $^3$D term to the $3d4s$ $^3$D term.

| $E_{upp}$ (cm$^{-1}$) | $J_{upp}$ | $E_{low}$ (cm$^{-1}$) | $J_{low}$ | $\lambda_{air}$ (Å) | BF This expt. | Unc. (%) | BF M17 | BF pure LS |
|---|---|---|---|---|---|---|---|---|
| 38192.392 | 1 | 20335.360 | 1 | 5598.4804 | 0.736 | 1 | 0.740 | 0.750 |
| 38192.392 | 1 | 20349.260 | 2 | 5602.8418 | 0.264 | 3 | 0.260 | 0.250 |
| 38219.118 | 2 | 20335.360 | 1 | 5590.1138 | 0.164 | 3 | 0.154 | 0.150 |
| 38219.118 | 2 | 20349.260 | 2 | 5594.4621 | 0.667 | 1 | 0.680 | 0.694 |
| 38219.118 | 2 | 20371.000 | 3 | 5601.2766 | 0.169 | 3 | 0.166 | 0.156 |
| 38259.124 | 3 | 20349.260 | 2 | 5581.9654 | 0.119 | 3 | 0.117 | 0.111 |
| 38259.124 | 3 | 20371.000 | 3 | 5588.7494 | 0.881 | 1 | 0.883 | 0.889 |

Table 5

Emission BF measurements on the multiplet from the upper 3*d*4*p* $^3$P term to the 3*d*4*s* $^3$D term.

| $E_{upp}$ ($cm^{-1}$) | $J_{upp}$ | $E_{low}$ ($cm^{-1}$) | $J_{low}$ | $\lambda_{air}$ (Å) | BF This expt. | Unc. (%) | BF M17 | BF pure LS |
|---|---|---|---|---|---|---|---|---|
| 39333.382 | 0 | 20335.360 | 1 | 5262.2413 | 1.000 | 0 | 0.992 | 1.000 |
| 39335.322 | 1 | 20335.360 | 1 | 5261.7040 | 0.251 | 3 | 0.259 | 0.250 |
| 39335.322 | 1 | 20349.260 | 2 | 5265.5562 | 0.749 | 1 | 0.732 | 0.750 |
| 39340.080 | 2 | 20335.360 | 1 | 5260.3867 | 0.009 | 20 | 0.011 | 0.010 |
| 39340.080 | 2 | 20349.260 | 2 | 5264.2370 | 0.174 | 6 | 0.160 | 0.150 |
| 39340.080 | 2 | 20371.000 | 3 | 5270.2702 | 0.791 | 2 | 0.820 | 0.840 |

Table 6

Emission BF measurements from the 3*d*4*p* $^1D_2$ level (first three rows) and from the 3*d*4*p* $^3$F term (last six rows) to the 3*d*4*s* $^1D_2$ level and the 3*d*4*s* $^3$D term[a]

| upper term | $E_{upp}$ ($cm^{-1}$) | lower term | $E_{low}$ ($cm^{-1}$) | $\lambda_{air}$ (Å) | BF This expt. | Unc. (%) | BF M17 | BF pure LS |
|---|---|---|---|---|---|---|---|---|
| $^1D_2$ | 35835.413 | $^3D_1$ | 20335.360 | 6449.8083 | 0.203 | 5 | 0.205 | 0.000 |
| $^1D_2$ | 35835.413 | $^3D_2$ | 20349.260 | 6455.5976 | 0.020 | 7 | 0.027 | 0.000 |
| $^1D_2$ | 35835.413 | $^1D_2$ | 21849.634 | 7148.1497 | 0.777 | 2 | 0.764 | 1.000 |
| $^3F_2$ | 35730.454 | $^3D_1$ | 20335.360 | 6493.7815 | 0.555 | 5 | 0.561 | 0.840 |
| $^3F_2$ | 35730.454 | $^3D_2$ | 20349.260 | 6499.6500 | 0.115 | 5 | 0.121 | 0.156 |
| $^3F_2$ | 35730.454 | $^1D_2$ | 21849.634 | 7202.2004 | 0.331 | 5 | 0.316 | 0.000 |
| $^3F_3$ | 35818.713 | $^3D_2$ | 20349.260 | 6462.5667 | 0.896 | 1 | 0.883 | 0.889 |
| $^3F_3$ | 35818.713 | $^3D_3$ | 20371.000 | 6471.6618 | 0.104 | 5 | 0.117 | 0.111 |
| $^3F_4$ | 35896.889 | $^3D_3$ | 20371.000 | 6439.0754 | 1.000 | 0 | 1.000 | 1.000 |

Note

[a] Although there is significant mixing between the $^1D_2$ and $^3F_2$ levels at 35835.413 $cm^{-1}$ and 35730.454 $cm^{-1}$

Table 7

*A*-values and log(*gf*)s for multiplets from the upper $4p^2$ $^3$P and $4s5s$ $^3S_1$ terms to the $4s4p$ $^3$P lower term.

| $\lambda_{air}$ (Å) | $E_{upp}$ (cm$^{-1}$) | $J_{upp}$ | $E_{low}$ (cm$^{-1}$) | $J_{low}$ | $A_{UW}$[a] ($10^6$ s$^{-1}$) | log(*gf*) UW | UW Unc. (%) | log(*gf*) M17 | log(*gf*) Ald(09)[a] | Unc. (%) |
|---|---|---|---|---|---|---|---|---|---|---|
| 4283.0106 | 38551.558 | 2 | 15210.063 | 1 | 44.4 | -0.215 | 6 | -0.201 | | |
| 4289.3668 | 38464.808 | 1 | 15157.901 | 0 | 60. | -0.306 | 5 | -0.296 | | |
| 4298.9883 | 38464.808 | 1 | 15210.063 | 1 | 45.8 | -0.419 | 5 | -0.414 | | |
| 4302.5278 | 38551.558 | 2 | 15315.943 | 2 | 132 | +0.264 | 5 | +0.281 | | |
| 4307.7439 | 38417.543 | 0 | 15210.063 | 1 | 172 | -0.319 | 5 | -0.294 | | |
| 4318.6517 | 38464.808 | 1 | 15315.943 | 2 | 72 | -0.219 | 5 | -0.198 | | |
| 6102.7227 | 31539.495 | 1 | 15157.901 | 0 | 9.3 | -0.809 | 6 | -0.810 | -0.85 | 9 |
| 6122.2172 | 31539.495 | 1 | 15210.063 | 1 | 27.3 | -0.337 | 5 | -0.332 | -0.38 | 9 |
| 6162.1730 | 31539.495 | 1 | 15315.943 | 2 | 44.8 | -0.116 | 5 | -0.109 | -0.17 | 9 |

Note:

[a] UW is this study; Ald(09) is Aldenius et al. (2009)

Table 8

Recommended transition probabilities for 202 lines of Ca I organized by increasing wavelength in air.

| $\lambda_{air}$[1] (Å) | Upper Level Energy (cm$^{-1}$) | Lower Level Energy (cm$^{-1}$) | Excitation Potential (eV) | M17 log(*gf*)[2] | Y&D[3] *L-V* % | Rec. log(*gf*) | Experimental Reference[4] | Exptl. Unc. % |
|---|---|---|---|---|---|---|---|---|
| 2200.7266 | 45425.358 | 0.000 | 0.000 | -1.549 | … | -1.49 | Park 76 | ±15 |
| 2223.6235 | 44957.655 | 0.000 | 0.000 | -9.411 | 87 | -9.411 | … | … |
| 2275.4656 | 43933.477 | 0.000 | 0.000 | -1.121 | -7 | -1.18 | Park 76 | ±15 |
| 2351.1863 | 42518.708 | 0.000 | 0.000 | -8.554 | -328 | -8.554 | … | … |
| 2398.5590 | 41679.008 | 0.000 | 0.000 | -1.624 | 11 | -1.36 | O&P61 | ±5 |
| 2541.4811 | 39335.322 | 0.000 | 0.000 | -4.610 | -7 | -4.610 | … | … |
| 2617.5413 | 38192.392 | 0.000 | 0.000 | -4.651 | -27 | -4.651 | … | … |

Notes:

The table is available in its entirety in the online version. Only a portion of the table is shown here to indicate its form and content.

[1] Air wavelengths are computed from NIST ASD energy levels and the index of refraction in air from Peck and Reeder (1972)

[2] The theoretical calculations of M17 do not have stated uncertainties.

[3] The % difference Length-Velocity forms of the calculation is taken from Yu & Derevianko (2018)

[4] Experimental references are coded: Park76 is Parkinson et al. (1976); O&P61 is Ostrovskii & Penkin (1961); UW+M17+Ueda83 are M17 log(*gf*)s rescaled using Ueda et al. (1983) and UW (this study); UW+M17+Ueda82 are M17 log(*gf*)s rescaled using Ueda et al. (1982) results that were renormalized using selected multiplets labeled M17+UW; M17+UW are M17 log(*gf*)s renormalized using UW lifetimes; Vogt07 is Vogt et al. (2007); S&R81 is Smith and Raggett (1981); Smith88 is Smith (1988). No entry indicates the recommended log(*gf*) is taken directly from M17.

Table 9

Ca I Equivalent Widths and Abundances

| λ[a] (Å) | χ[a] (eV) | log(*gf*)[a] | EW[b] (mÅ) Sun | EW[b] (mÅ) HD 84937 | log ε[c] Sun | log ε[c] HD 84937 |
|---|---|---|---|---|---|---|
| 2275.466 | 0.000 | -1.180 | ... | ... | ... | 4.46 |
| 2398.559 | 0.000 | -1.360 | ... | ... | ... | 4.41 |
| 3644.413 | 1.899 | -0.281 | ... | 19.6 | ... | 4.39 |
| 3875.776 | 2.526 | -0.842 | ... | 4.0 | ... | 4.56 |
| 4094.925 | 2.523 | -0.686 | ... | 4.0 | ... | 4.39 |
| 4226.728 | 0.000 | 0.239 | ... | 153.0 | ... | 4.48 |
| 4283.011 | 1.886 | -0.215 | ... | 35.7 | ... | 4.49 |
| 4289.367 | 1.879 | -0.310 | ... | 29.5 | ... | 4.45 |
| 4298.988 | 1.886 | -0.428 | ... | 26.0 | ... | 4.49 |
| 4302.528 | 1.899 | 0.267 | ... | 60.0 | ... | 4.48 |
| 4318.652 | 1.899 | -0.212 | ... | 34.5 | ... | 4.47 |
| 4355.079 | 2.709 | -0.426 | ... | 4.9 | ... | 4.37 |
| 4425.437 | 1.879 | -0.393 | ... | 27.6 | ... | 4.48 |
| 4434.957 | 1.886 | -0.040 | ... | 44.3 | ... | 4.46 |
| 4435.679 | 1.886 | -0.528 | ... | 21.5 | ... | 4.47 |
| 4454.779 | 1.899 | 0.228 | ... | 59.5 | ... | 4.47 |
| 4455.887 | 1.899 | -0.536 | ... | 21.1 | ... | 4.48 |
| 4456.616 | 1.899 | -1.723 | ... | 2.1 | ... | 4.57 |
| 4509.447 | 2.523 | -1.891 | 17.5 | ... | 6.20 | ... |
| 4512.268 | 2.526 | -1.900 | 23.2 | ... | 6.40 | ... |
| 4526.928 | 2.709 | -0.548 | ... | 5.5 | ... | 4.54 |
| 4578.551 | 2.521 | -0.697 | 80.7 | 4.5 | 6.28 | 4.43 |
| 4581.395 | 2.523 | -0.502 | ... | 9.6 | ... | 4.59 |
| 4585.866 | 2.526 | -0.338 | ... | 11.0 | ... | 4.49 |
| 5260.387 | 2.521 | -1.719 | 29.4 | ... | 6.33 | ... |
| 5261.704 | 2.521 | -0.579 | 95.0 | 6.7 | 6.42 | 4.47 |
| 5265.556 | 2.523 | -0.113 | ... | 16.0 | ... | 4.43 |
| 5349.465 | 2.709 | -0.310 | 98.3 | 8.3 | 6.27 | 4.47 |
| 5512.980 | 2.933 | -0.464 | 91.0 | 4.5 | 6.47 | 4.53 |
| 5581.965 | 2.523 | -0.555 | 92.0 | 7.5 | 6.31 | 4.49 |
| 5588.749 | 2.526 | 0.358 | 159.1 | 34.8 | 6.22 | 4.42 |
| 5590.114 | 2.521 | -0.571 | 89.7 | 6.8 | 6.29 | 4.46 |
| 5594.462 | 2.523 | 0.097 | 133.9 | 23.8 | 6.24 | 4.44 |
| 5598.480 | 2.521 | -0.087 | ... | 18.0 | ... | 4.46 |
| 5601.277 | 2.526 | -0.523 | ... | 7.3 | ... | 4.45 |
| 5857.451 | 2.933 | 0.240 | 146.0 | 17.0 | 6.39 | 4.46 |
| 5867.562 | 2.933 | -1.570 | 24.3 | ... | 6.40 | ... |
| 6097.266 | 2.521 | -3.177 | 2.3 | ... | 6.48 | ... |
| 6102.723 | 1.879 | -0.810 | 147.0 | 16.0 | 6.37 | 4.53 |
| 6122.217 | 1.886 | -0.339 | 214.0 | 33.5 | 6.33 | 4.49 |
| 6161.297 | 2.523 | -1.266 | 57.1 | ... | 6.29 | ... |
| 6162.173 | 1.899 | -0.116 | 270.0 | 44.5 | 6.37 | 4.48 |
| 6163.755 | 2.521 | -1.286 | 64.2 | ... | 6.41 | ... |
| 6166.439 | 2.521 | -1.142 | 72.0 | 2.2 | 6.39 | 4.50 |
| 6169.042 | 2.523 | -0.797 | 99.1 | 5.5 | 6.42 | 4.57 |
| 6169.563 | 2.526 | -0.478 | 116.8 | 8.4 | 6.31 | 4.46 |
| 6439.075 | 2.526 | 0.390 | 173.7 | 38.7 | 6.23 | 4.45 |
| 6449.808 | 2.521 | -0.502 | 101.5 | 10.0 | 6.32 | 4.55 |
| 6455.598 | 2.523 | -1.340 | 54.1 | ... | 6.35 | ... |
| 6462.567 | 2.523 | 0.262 | ... | 31.3 | ... | 4.42 |
| 6464.673 | 2.526 | -2.249 | 11.3 | ... | 6.27 | ... |
| 6471.662 | 2.526 | -0.686 | 93.5 | 5.9 | 6.39 | 4.49 |
| 6493.782 | 2.521 | -0.109 | 129.5 | 20.0 | 6.30 | 4.52 |
| 6499.650 | 2.523 | -0.818 | 88.4 | 5.7 | 6.42 | 4.60 |
| 6508.850 | 2.526 | -2.618 | 8.1 | ... | 6.48 | ... |
| 6572.779 | 0.000 | -4.320 | 26.8 | ... | 6.38 | ... |
| 6709.893 | 2.933 | -2.879 | 2.1 | ... | 6.50 | ... |
| 6717.681 | 2.709 | -0.524 | ... | 6.4 | ... | 4.52 |
| 6798.479 | 2.709 | -2.660 | 5.4 | ... | 6.49 | ... |
| 7148.150 | 2.709 | 0.137 | 152.1 | 22.5 | 6.33 | 4.49 |
| 7202.200 | 2.709 | -0.262 | 117.4 | 12.9 | 6.36 | 4.58 |
| 7326.146 | 2.933 | -0.208 | 115.0 | 9.0 | 6.44 | 4.55 |
| 9099.089 | 3.910 | -1.281 | 6.6 | ... | 6.19 | ... |
| 9108.827 | 3.910 | -2.005 | 1.2 | ... | 6.15 | ... |
| 9416.971 | 4.131 | -0.808 | 10.9 | ... | 6.13 | ... |

Notes:

[a] Ritz wavelengths and excitation potentials and (recommended) log(*gf*)s are taken from Table 8

[b] Equivalent widths; typical uncertainties are ~0.5 mÅ for the Sun and ~1.0 mÅ for HD 84937

[c] log10 ε(X) = log10 (NX/NH) + 12; uncertainties from EW measurements are typically ±0.03 for individual lines; rising to ±0.04 for strong lines

Table 10

Ca II Abundances

| λ[a] (Å) | χ[a] (eV) | log(*gf*) | log ε Sun | log ε HD 84937 |
|---|---|---|---|---|
| 2113.146 | 3.151 | -1.369 | ... | 4.56 |
| 2128.750 | 1.692 | -3.319 | ... | 4.46 |
| 2131.505 | 1.700 | -2.371 | ... | 4.51 |
| 2132.304 | 1.692 | -2.658 | ... | 4.41 |
| 2197.787 | 3.123 | -1.330 | ... | 4.66 |
| 2208.611 | 3.151 | -1.027 | ... | 4.46 |
| 3158.869 | 3.123 | 0.246 | ... | 4.54 |
| 3179.331 | 3.151 | 0.504 | ... | 4.56 |
| 3181.275 | 3.151 | -0.450 | ... | 4.47 |
| 3706.024 | 3.123 | -0.453 | ... | 4.48 |
| 3736.902 | 3.151 | -0.146 | ... | 4.44 |
| 3933.663 | 0.000 | 0.113 | ... | 4.56 |
| 5001.479 | 7.505 | -0.521 | 6.31 | ... |
| 5021.138 | 7.515 | -1.222 | 6.34 | ... |
| 5285.266 | 7.505 | -1.153 | 6.37 | ... |
| 8201.72 | 7.505 | 0.318 | 6.31 | ... |
| 8248.80 | 7.515 | 0.576 | 6.34 | ... |
| 8254.72 | 7.515 | -0.378 | 6.35 | ... |

Notes:

[a] Ritz wavelengths and excitation energies taken from the NIST ASD (Kramida et al. 2019)

[b] log(*gf*)s are taken from the theoretical calculations of Theodosiou (1989) for the first four entries and from Safranova & Safranova (2011) for the remaining entries

Title: ATOMIC TRANSITION PROBABILILTIES OF NEUTRAL CALCIUM

Authors: E. A. Den Hartog, J. E. Lawler, C. Sneden, J. J. Cowan, I. U. Roederer & J. Sobeck

Table 8. Recommended transition probabilities for 202 lines of Ca I organized by increasing wavelength in air.

================================================================================

Byte-by-byte Description of file: apjsxxxxt8_mrt.txt

| Bytes | Format | Units | Label | Explanations |
|---|---|---|---|---|
| 1- 9 | F9.4 | Angstrom | Wave | Wavelength in air (1) |
| 11- 19 | F9.3 | cm^-1 | E_upper | Energy of Upper Level |
| 21- 29 | F9.3 | cm^-1 | E_Lower | Energy of Lower Level |
| 31- 35 | F5.3 | eV | ExcPot | Excitation Potential |
| 37- 43 | F7.3 | none | log(gf)_M17 | logarithm of the degeneracy multiplied by the oscillator strength from M17 (2) |
| 45- 50 | F5.0 | % | L-V_Y&D | percent difference between length and velocity forms of the calculations (3) |
| 52- 59 | F8.5 | none | log(gf)_rec | recommended logarithm of the degeneracy multiplied by the oscillator strength |
| 61- 74 | A14 | none | ExptRef | experimental reference (4) |
| 76- 82 | F5.2 | % | ExptUnc | percent experimental uncertainty |

Note (1): Air wavelengths are computed from NIST ASD energy levels and the index of refraction in air from Peck and Reeder (1972)

(2): The theoretical calculations of M17 do not have stated uncertainties

(3): The % difference Length-Velocity forms of the calculations taken from Yu & Derevianko (2018)

(4): Experimental references are coded:
Park76 is Parkinson et al. (1976);
O&P61 is Otrovskii & Penkin (1961);
UW+M17+Ueda83 are M17 log(gf)s rescaled using Ueda et al. (1983) and UW (this study);
UW+M17+Ueda82 are M17 log(gf)s rescaled using Ueda et al. (1982) results that were renormalized using selected multiplets labeled M17+UW;
M17+UW are M17 log(gf)s renormalized using UW lifetimes;
Vogt07 is Vogt et al. (2007);
S&R81 is Smith and Raggett (1981);
Smith88 is Smith (1988).
No entry indicates the recommended log(gf) is taken directly from M17

| | | | | | | | | |
|---|---|---|---|---|---|---|---|---|
| 2200.7266 | 45425.358 | 0.000 | 0.000 | -1.549 | | -1.49 | Park76 | 15 |
| 2223.6235 | 44957.655 | 0.000 | 0.000 | -9.411 | 87 | -9.411 | | |
| 2275.4656 | 43933.477 | 0.000 | 0.000 | -1.121 | -7 | -1.18 | Park76 | 15 |
| 2351.1863 | 42518.708 | 0.000 | 0.000 | -8.554 | -328 | -8.554 | | |
| 2398.5590 | 41679.008 | 0.000 | 0.000 | -1.624 | 11 | -1.36 | O&P61 | 5 |
| 2541.4811 | 39335.322 | 0.000 | 0.000 | -4.610 | -7 | -4.610 | | |
| 2617.5413 | 38192.392 | 0.000 | 0.000 | -4.651 | -27 | -4.651 | | |
| 2721.6442 | 36731.615 | 0.000 | 0.000 | -2.602 | -144 | -3.05 | O&P61 | 5 |
| 2734.8132 | 36554.749 | 0.000 | 0.000 | -6.104 | -1791 | -6.104 | | |
| 3258.8169 | 45887.200 | 15210.063 | 1.886 | -8.011 | | -8.011 | | |
| 3269.0851 | 45738.684 | 15157.901 | 1.879 | -2.447 | | -2.447 | | |
| 3274.6709 | 45738.684 | 15210.063 | 1.886 | -1.969 | | -1.969 | | |
| 3286.0681 | 45738.684 | 15315.943 | 1.899 | -1.746 | | -1.746 | | |
| 3344.5078 | 45049.073 | 15157.901 | 1.879 | -1.319 | -3 | -1.318 | UW+M17+Ueda83 | 24 |
| 3350.2035 | 45050.419 | 15210.063 | 1.886 | -0.967 | -3 | -0.966 | UW+M17+Ueda83 | 17 |
| 3350.3546 | 45049.073 | 15210.063 | 1.886 | -1.443 | -3 | -1.442 | UW+M17+Ueda83 | 17 |
| 3357.0199 | 44989.830 | 15210.063 | 1.886 | -3.823 | -3 | -3.823 | | |
| 3361.9124 | 45052.374 | 15315.943 | 1.899 | -0.441 | -3 | -0.440 | UW+M17+Ueda83 | 11 |
| 3362.1334 | 45050.419 | 15315.943 | 1.899 | -1.442 | -3 | -1.441 | UW+M17+Ueda83 | 11 |
| 3362.2856 | 45049.073 | 15315.943 | 1.899 | -2.617 | -3 | -2.616 | UW+M17+Ueda83 | 11 |
| 3368.9985 | 44989.830 | 15315.943 | 1.899 | -4.322 | -3 | -4.322 | | |
| 3439.4039 | 44276.538 | 15210.063 | 1.886 | -6.908 | -3 | -6.908 | | |
| 3468.4744 | 43980.767 | 15157.901 | 1.879 | -2.184 | 5 | -2.184 | | |
| 3474.7630 | 43980.767 | 15210.063 | 1.886 | -1.705 | 5 | -1.705 | | |
| 3487.5982 | 43980.767 | 15315.943 | 1.899 | -1.479 | 5 | -1.479 | | |
| 3536.9631 | 43474.827 | 15210.063 | 1.886 | -7.041 | 69 | -7.041 | | |
| 3548.4618 | 43489.119 | 15315.943 | 1.899 | -6.679 | 62 | -6.679 | | |
| 3550.2628 | 43474.827 | 15315.943 | 1.899 | -8.577 | 96 | -8.577 | | |
| 3607.9079 | 42919.053 | 15210.063 | 1.886 | -5.312 | -10 | -5.312 | | |
| 3621.7475 | 42919.053 | 15315.943 | 1.899 | -5.044 | -15 | -5.044 | | |
| 3624.1120 | 42743.002 | 15157.901 | 1.879 | -0.944 | -1 | -0.907 | UW+M17+Ueda82 | 13 |
| 3630.7522 | 42744.716 | 15210.063 | 1.886 | -0.591 | -1 | -0.554 | UW+M17+Ueda82 | 11 |
| 3630.9782 | 42743.002 | 15210.063 | 1.886 | -1.065 | -1 | -1.028 | UW+M17+Ueda82 | 11 |
| 3644.4130 | 42747.387 | 15315.943 | 1.899 | -0.318 | -1 | -0.281 | UW+M17+Ueda82 | 12 |
| 3644.7679 | 42744.716 | 15315.943 | 1.899 | -1.060 | -1 | -1.023 | UW+M17+Ueda82 | 12 |
| 3644.9957 | 42743.002 | 15315.943 | 1.899 | -2.233 | -1 | -2.196 | UW+M17+Ueda82 | 12 |
| 3761.6944 | 41786.276 | 15210.063 | 1.886 | -3.638 | -8 | -3.638 | | |
| 3869.8998 | 46182.399 | 20349.260 | 2.523 | -3.846 | | -3.846 | | |
| 3870.4774 | 46164.644 | 20335.360 | 2.521 | -1.176 | | -1.176 | | |
| 3872.5403 | 46164.785 | 20349.260 | 2.523 | -1.005 | | -1.005 | | |
| 3872.5615 | 46164.644 | 20349.260 | 2.523 | -1.907 | | -1.907 | | |
| 3873.1594 | 46182.399 | 20371.000 | 2.526 | -6.942 | | -6.942 | | |
| 3875.7764 | 46164.971 | 20371.000 | 2.526 | -0.842 | | -0.842 | | |
| 3875.8043 | 46164.785 | 20371.000 | 2.526 | -1.905 | | -1.905 | | |
| 3875.8255 | 46164.644 | 20371.000 | 2.526 | -3.448 | | -3.448 | | |
| 3918.9547 | 40719.847 | 15210.063 | 1.886 | -4.277 | -16 | -4.277 | | |
| 3923.4785 | 40690.435 | 15210.063 | 1.886 | -3.751 | -9 | -3.751 | | |
| 3935.2888 | 40719.847 | 15315.943 | 1.899 | -3.040 | -10 | -3.040 | | |
| 3948.9004 | 40474.241 | 15157.901 | 1.879 | -1.735 | 5 | -1.661 | UW+M17+Ueda82 | 16 |
| 3957.0537 | 40474.241 | 15210.063 | 1.886 | -1.249 | 4 | -1.175 | UW+M17+Ueda82 | 15 |
| 3973.7076 | 40474.241 | 15315.943 | 1.899 | -1.010 | 4 | -0.936 | UW+M17+Ueda82 | 15 |
| 3984.5250 | 45425.358 | 20335.360 | 2.521 | -7.810 | -1504 | -7.810 | | |
| 3986.7338 | 45425.358 | 20349.260 | 2.523 | -4.529 | -22 | -4.529 | | |
| 4059.5368 | 44961.757 | 20335.360 | 2.521 | -5.013 | -131 | -5.013 | | |
| 4060.2131 | 44957.655 | 20335.360 | 2.521 | -3.865 | -137 | -3.865 | | |
| 4060.5405 | 44955.670 | 20335.360 | 2.521 | -3.755 | -140 | -3.755 | | |
| 4061.8295 | 44961.757 | 20349.260 | 2.523 | -3.840 | -133 | -3.840 | | |
| 4062.5066 | 44957.655 | 20349.260 | 2.523 | -3.390 | -140 | -3.390 | | |
| 4065.4205 | 44961.757 | 20371.000 | 2.526 | -3.096 | -137 | -3.096 | | |
| 4087.8860 | 44804.878 | 20349.260 | 2.523 | -4.206 | 0 | -4.206 | | |
| 4091.5233 | 44804.878 | 20371.000 | 2.526 | -6.220 | -16 | -6.220 | | |
| 4092.6318 | 44762.620 | 20335.360 | 2.521 | -0.857 | -2 | -0.857 | | |
| 4094.9253 | 44762.839 | 20349.260 | 2.523 | -0.686 | -2 | -0.686 | | |
| 4094.9621 | 44762.620 | 20349.260 | 2.523 | -1.587 | -2 | -1.587 | | |
| 4098.5282 | 44763.118 | 20371.000 | 2.526 | -0.523 | -2 | -0.523 | | |
| 4098.5751 | 44762.839 | 20371.000 | 2.526 | -1.586 | -2 | -1.586 | | |
| 4098.6119 | 44762.620 | 20371.000 | 2.526 | -3.130 | -2 | -3.130 | | |
| 4108.5257 | 46182.399 | 21849.634 | 2.709 | -0.828 | | -0.828 | | |
| 4111.5020 | 46164.785 | 21849.634 | 2.709 | -3.872 | | -3.872 | | |
| 4111.5258 | 46164.644 | 21849.634 | 2.709 | -4.869 | | -4.869 | | |
| 4226.7276 | 23652.304 | 0.000 | 0.000 | 0.242 | -7 | 0.23884 | Vogt07 | 0.04 |
| 4236.4333 | 43933.477 | 20335.360 | 2.521 | -6.217 | 63 | -6.217 | | |
| 4238.9302 | 43933.477 | 20349.260 | 2.523 | -5.339 | -5 | -5.339 | | |
| 4240.4573 | 45425.358 | 21849.634 | 2.709 | -1.232 | -23 | -1.232 | | |
| 4283.0106 | 38551.558 | 15210.063 | 1.886 | -0.201 | -11 | -0.215 | UW+M17+Ueda82 | 5 |
| 4289.3668 | 38464.808 | 15157.901 | 1.879 | -0.296 | -11 | -0.310 | UW+M17+Ueda82 | 5 |
| 4298.9883 | 38464.808 | 15210.063 | 1.886 | -0.414 | -11 | -0.428 | UW+M17+Ueda82 | 5 |
| 4302.5278 | 38551.558 | 15315.943 | 1.899 | 0.281 | -11 | 0.267 | UW+M17+Ueda82 | 5 |
| 4307.7439 | 38417.543 | 15210.063 | 1.886 | -0.294 | -11 | -0.308 | UW+M17+Ueda82 | 5 |
| 4318.6517 | 38464.808 | 15315.943 | 1.899 | -0.198 | -11 | -0.212 | UW+M17+Ueda82 | 5 |
| 4325.5173 | 44961.757 | 21849.634 | 2.709 | -6.827 | -196 | -6.827 | | |
| 4326.2852 | 44957.655 | 21849.634 | 2.709 | -6.035 | -5 | -6.035 | | |
| 4355.0790 | 44804.878 | 21849.634 | 2.709 | -0.426 | 1 | -0.426 | | |
| 4363.0694 | 44762.839 | 21849.634 | 2.709 | -4.183 | 0 | -4.183 | | |
| 4363.1111 | 44762.620 | 21849.634 | 2.709 | -5.034 | -2 | -5.034 | | |
| 4425.4368 | 37748.197 | 15157.901 | 1.879 | -0.410 | -1 | -0.393 | M17+UW | 5 |
| 4434.9570 | 37751.867 | 15210.063 | 1.886 | -0.057 | -1 | -0.040 | M17+UW | 5 |
| 4435.6792 | 37748.197 | 15210.063 | 1.886 | -0.545 | -1 | -0.528 | M17+UW | 5 |
| 4454.7785 | 37757.449 | 15315.943 | 1.899 | 0.211 | -1 | 0.228 | M17+UW | 5 |
| 4455.8869 | 37751.867 | 15315.943 | 1.899 | -0.553 | 0 | -0.536 | M17+UW | 5 |
| 4456.6159 | 37748.197 | 15315.943 | 1.899 | -1.740 | 0 | -1.723 | M17+UW | 5 |
| 4496.1737 | 45887.200 | 23652.304 | 2.933 | -1.966 | | -1.966 | | |
| 4505.0208 | 42526.591 | 20335.360 | 2.521 | -3.514 | -57 | -3.514 | | |
| 4506.6218 | 42518.708 | 20335.360 | 2.521 | -2.357 | -59 | -2.357 | | |
| 4507.4067 | 42514.845 | 20335.360 | 2.521 | -2.242 | -59 | -2.242 | | |
| 4507.8445 | 42526.591 | 20349.260 | 2.523 | -2.348 | -58 | -2.348 | | |
| 4509.4474 | 42518.708 | 20349.260 | 2.523 | -1.891 | -60 | -1.891 | | |
| 4512.2678 | 42526.591 | 20371.000 | 2.526 | -1.616 | -60 | -1.900 | S&R81 | 7 |
| 4526.0301 | 37298.287 | 15210.063 | 1.886 | -5.207 | 42 | -5.207 | | |
| 4526.4080 | 45738.684 | 23652.304 | 2.933 | -6.627 | | -6.627 | | |
| 4526.9280 | 43933.477 | 21849.634 | 2.709 | -0.470 | -21 | -0.548 | S&R81 | 5 |
| 4545.3527 | 42343.587 | 20349.260 | 2.523 | -3.936 | 8 | -3.936 | | |
| 4547.8305 | 37298.287 | 15315.943 | 1.899 | -3.385 | -7 | -3.385 | | |
| 4549.8500 | 42343.587 | 20371.000 | 2.526 | -4.454 | -9 | -4.454 | | |
| 4578.5506 | 42170.214 | 20335.360 | 2.521 | -0.674 | 0 | -0.697 | S&R81 | 5 |
| 4581.3950 | 42170.558 | 20349.260 | 2.523 | -0.502 | 0 | -0.502 | | |
| 4581.4672 | 42170.214 | 20349.260 | 2.523 | -1.404 | 0 | -1.404 | | |
| 4585.8655 | 42171.026 | 20371.000 | 2.526 | -0.338 | 0 | -0.338 | | |
| 4585.9639 | 42170.558 | 20371.000 | 2.526 | -1.403 | 0 | -1.403 | | |
| 4586.0363 | 42170.214 | 20371.000 | 2.526 | -2.948 | 0 | -2.948 | | |
| 4672.0011 | 45050.419 | 23652.304 | 2.933 | -3.771 | -6 | -3.771 | | |
| 4672.2950 | 45049.073 | 23652.304 | 2.933 | -5.843 | -12 | -5.843 | | |
| 4683.9238 | 41679.008 | 20335.360 | 2.521 | -6.197 | 77 | -6.197 | | |
| 4685.2677 | 44989.830 | 23652.304 | 2.933 | -1.023 | | -0.879 | Smith88 | 6 |
| 4686.9762 | 41679.008 | 20349.260 | 2.523 | -4.688 | -29 | -4.688 | | |
| 4834.9506 | 42526.591 | 21849.634 | 2.709 | -5.216 | -38 | -5.216 | | |
| 4836.7947 | 42518.708 | 21849.634 | 2.709 | -5.590 | 12 | -5.590 | | |
| 4847.3107 | 44276.538 | 23652.304 | 2.933 | -1.618 | | -1.400 | Smith88 | 7 |
| 4878.1257 | 42343.587 | 21849.634 | 2.709 | -0.109 | 6 | -0.164 | S&R81 | 5 |
| 4917.8381 | 43980.767 | 23652.304 | 2.933 | -6.334 | -5 | -6.334 | | |
| 4919.6626 | 42170.558 | 21849.634 | 2.709 | -4.064 | 1 | -4.064 | | |
| 4919.7459 | 42170.214 | 21849.634 | 2.709 | -4.778 | -1 | -4.778 | | |
| 4951.9004 | 40537.893 | 20349.260 | 2.523 | -4.652 | 41 | -4.652 | | |
| 4957.2386 | 40537.893 | 20371.000 | 2.526 | -3.490 | -8 | -3.490 | | |
| 5041.6177 | 41679.008 | 21849.634 | 2.709 | -0.565 | -15 | -0.471 | S&R81 | 6 |
| 5043.3602 | 43474.827 | 23652.304 | 2.933 | -4.124 | -15 | -4.124 | | |
| 5188.8442 | 42919.053 | 23652.304 | 2.933 | -0.022 | -5 | -0.075 | Smith88 | 5 |
| 5236.2253 | 42744.716 | 23652.304 | 2.933 | -4.331 | -6 | -4.331 | | |
| 5236.6954 | 42743.002 | 23652.304 | 2.933 | -5.239 | -8 | -5.239 | | |
| 5260.3867 | 39340.080 | 20335.360 | 2.521 | -1.737 | -23 | -1.719 | S&R81 | 5 |
| 5261.7040 | 39335.322 | 20335.360 | 2.521 | -0.598 | -23 | -0.579 | S&R81 | 5 |
| 5262.2413 | 39333.382 | 20335.360 | 2.521 | -0.492 | -23 | -0.471 | S&R81 | 5 |
| 5264.2370 | 39340.080 | 20349.260 | 2.523 | -0.586 | -23 | -0.574 | S&R81 | 5 |
| 5265.5562 | 39335.322 | 20349.260 | 2.523 | -0.147 | -24 | -0.113 | S&R81 | 5 |
| 5270.2702 | 39340.080 | 20371.000 | 2.526 | 0.124 | -24 | 0.162 | S&R81 | 5 |
| 5349.4654 | 40537.893 | 21849.634 | 2.709 | -0.766 | 24 | -0.310 | S&R81 | 5 |
| 5512.9802 | 41786.276 | 23652.304 | 2.933 | -0.450 | -12 | -0.464 | Smith88 | 5 |
| 5521.1311 | 33317.264 | 15210.063 | 1.886 | -5.111 | -8 | -5.111 | | |
| 5581.9654 | 38259.124 | 20349.260 | 2.523 | -0.578 | -11 | -0.555 | S&R81 | 5 |
| 5588.7494 | 38259.124 | 20371.000 | 2.526 | 0.302 | -12 | 0.358 | S&R81 | 5 |
| 5590.1138 | 38219.118 | 20335.360 | 2.521 | -0.604 | -11 | -0.571 | S&R81 | 5 |
| 5594.4621 | 38219.118 | 20349.260 | 2.523 | 0.042 | -12 | 0.097 | S&R81 | 5 |
| 5598.4804 | 38192.392 | 20335.360 | 2.521 | -0.143 | -11 | -0.087 | S&R81 | 5 |
| 5601.2766 | 38219.118 | 20371.000 | 2.526 | -0.569 | -12 | -0.523 | S&R81 | 5 |
| 5602.8418 | 38192.392 | 20349.260 | 2.523 | -0.598 | -12 | -0.564 | S&R81 | 5 |
| 5715.8214 | 39340.080 | 21849.634 | 2.709 | -3.238 | -14 | -3.238 | | |
| 5717.3767 | 39335.322 | 21849.634 | 2.709 | -4.326 | -24 | -4.326 | | |
| 5857.4505 | 40719.847 | 23652.304 | 2.933 | 0.171 | -10 | 0.240 | Smith88 | 5 |
| 5867.5620 | 40690.435 | 23652.304 | 2.933 | -2.135 | -54 | -1.570 | Smith88 | 9 |
| 5942.9721 | 40474.241 | 23652.304 | 2.933 | -5.730 | -5 | -5.730 | | |
| 6092.3481 | 38259.124 | 21849.634 | 2.709 | -6.162 | -373 | -6.162 | | |
| 6097.2658 | 36731.615 | 20335.360 | 2.521 | -3.177 | -7 | -3.177 | | |
| 6102.4392 | 36731.615 | 20349.260 | 2.523 | -3.404 | 2 | -3.404 | | |
| 6102.7227 | 31539.495 | 15157.901 | 1.879 | -0.810 | 3 | -0.817 | M17+UW | 5 |
| 6107.2375 | 38219.118 | 21849.634 | 2.709 | -3.396 | -17 | -3.396 | | |
| 6117.2250 | 38192.392 | 21849.634 | 2.709 | -4.479 | 16 | -4.479 | | |
| 6122.2172 | 31539.495 | 15210.063 | 1.886 | -0.332 | 3 | -0.339 | M17+UW | 5 |
| 6156.0232 | 36575.119 | 20335.360 | 2.521 | -2.749 | -9 | -2.749 | | |
| 6161.2969 | 36575.119 | 20349.260 | 2.523 | -1.353 | -5 | -1.266 | S&R81 | 6 |
| 6162.1730 | 31539.495 | 15315.943 | 1.899 | -0.109 | 3 | -0.116 | M17+UW | 5 |
| 6163.7547 | 36554.749 | 20335.360 | 2.521 | -1.364 | -5 | -1.286 | S&R81 | 5 |
| 6166.4392 | 36547.688 | 20335.360 | 2.521 | -1.225 | -5 | -1.142 | S&R81 | 5 |
| 6169.0416 | 36554.749 | 20349.260 | 2.523 | -0.871 | -6 | -0.797 | S&R81 | 5 |
| 6169.5631 | 36575.119 | 20371.000 | 2.526 | -0.604 | -6 | -0.478 | S&R81 | 5 |
| 6439.0754 | 35896.889 | 20371.000 | 2.526 | 0.333 | 8 | 0.390 | S&R81 | 5 |
| 6449.8083 | 35835.413 | 20335.360 | 2.521 | -0.546 | 8 | -0.502 | S&R81 | 5 |
| 6455.5976 | 35835.413 | 20349.260 | 2.523 | -1.422 | 11 | -1.340 | Smith88 | 9 |
| 6462.5667 | 35818.713 | 20349.260 | 2.523 | 0.169 | 8 | 0.262 | S&R81 | 5 |
| 6464.6730 | 35835.413 | 20371.000 | 2.526 | -2.249 | -3 | -2.249 | | |
| 6471.6618 | 35818.713 | 20371.000 | 2.526 | -0.709 | 8 | -0.686 | S&R81 | 5 |
| 6493.7815 | 35730.454 | 20335.360 | 2.521 | -0.148 | 8 | -0.109 | S&R81 | 5 |
| 6499.6500 | 35730.454 | 20349.260 | 2.523 | -0.814 | 7 | -0.818 | S&R81 | 5 |
| 6508.8497 | 35730.454 | 20371.000 | 2.526 | -2.618 | 15 | -2.618 | | |
| 6572.7792 | 15210.063 | 0.000 | 0.000 | -4.274 | -29 | -4.32 | Park76 | 26 |
| 6709.8930 | 38551.558 | 23652.304 | 2.933 | -2.879 | -6 | -2.879 | | |
| 6717.6810 | 36731.615 | 21849.634 | 2.709 | -0.578 | -11 | -0.524 | S&R81 | 5 |
| 6749.1900 | 38464.808 | 23652.304 | 2.933 | -5.015 | -28 | -5.015 | | |
| 6770.7950 | 38417.543 | 23652.304 | 2.933 | -3.687 | -17 | -3.687 | | |
| 6789.0741 | 36575.119 | 21849.634 | 2.709 | -2.208 | -12 | -2.208 | | |
| 6798.4786 | 36554.749 | 21849.634 | 2.709 | -2.660 | -12 | -2.660 | | |
| 6835.6509 | 46164.644 | 31539.495 | 3.910 | -9.965 | | -9.965 | | |
| 7090.4635 | 37751.867 | 23652.304 | 2.933 | -3.721 | 3 | -3.721 | | |
| 7092.3096 | 37748.197 | 23652.304 | 2.933 | -4.477 | -5 | -4.477 | | |
| 7148.1497 | 35835.413 | 21849.634 | 2.709 | 0.116 | -13 | 0.137 | S&R81 | 5 |
| 7156.6954 | 35818.713 | 21849.634 | 2.709 | -3.567 | 11 | -3.567 | | |
| 7199.5847 | 45425.358 | 31539.495 | 3.910 | -5.555 | 29 | -5.555 | | |
| 7202.2004 | 35730.454 | 21849.634 | 2.709 | -0.307 | -13 | -0.262 | S&R81 | 5 |
| 7326.1455 | 37298.287 | 23652.304 | 2.933 | -0.191 | 5 | -0.208 | Smith88 | 5 |
| 7448.2577 | 44961.757 | 31539.495 | 3.910 | -2.329 | -8 | -2.329 | | |
| 7450.5347 | 44957.655 | 31539.495 | 3.910 | -2.592 | -8 | -2.592 | | |
| 7451.6371 | 44955.670 | 31539.495 | 3.910 | -3.090 | -8 | -3.090 | | |
| 7560.4273 | 44762.620 | 31539.495 | 3.910 | -11.761 | 41 | -11.761 | | |
| 8066.2138 | 43933.477 | 31539.495 | 3.910 | -6.236 | 30 | -6.236 | | |
| 8256.6686 | 45425.358 | 33317.264 | 4.131 | -1.512 | -6 | -1.512 | | |
| 8588.4174 | 44957.655 | 33317.264 | 4.131 | -7.421 | 53 | -7.421 | | |
| 9099.0886 | 42526.591 | 31539.495 | 3.910 | -1.281 | -1 | -1.281 | | |
| 9105.6217 | 42518.708 | 31539.495 | 3.910 | -1.519 | -1 | -1.519 | | |
| 9108.8266 | 42514.845 | 31539.495 | 3.910 | -2.005 | -1 | -2.005 | | |
| 9404.1214 | 42170.214 | 31539.495 | 3.910 | -11.731 | 52 | -11.731 | | |
| 9416.9712 | 43933.477 | 33317.264 | 4.131 | -0.808 | -3 | -0.808 | | |
| 9859.7031 | 41679.008 | 31539.495 | 3.910 | -6.611 | -4 | -6.611 | | |
| 9989.0382 | 45738.684 | 35730.454 | 4.430 | -7.459 | | -7.459 | | |